\documentclass{ar2e}

\usepackage{natbib}
\usepackage{graphicx}
\usepackage{url}

\bibpunct{(}{)}{;}{a}{}{,}

\begin{document}

\input epsf.tex    
\input epsf.def   

\input psfig.sty

\def\nat{{\em Nature }}
\def\aap{{\em Astron. \& Astrophys. }}
\def\aj{{\em Astron.~J. }}
\def\apj{{\em Astrophys.~J. }}
\def\araa{{\em Ann. Rev. Astron. Astrophys. }}
\def\apjl{{\em Astrophys.~J.~Letters }}
\def\apjs{{\em Astrophys.~J.~Suppl. }}
\def\apss{{\em Astrophys.~Space~Sci. }}
\def\icarus{{\em Icarus }}
\def\mnras{{\em MNRAS }}
\def\pasp{{\em Pub. Astron. Soc. Pacific }}
\def\planss{{\em Plan. Space Sci. }}
\def\physrep{{\em Phys. Rep.}}
\def\bain{{\em Bull.~Astron.~Inst.~Netherlands }}
\def\lesssim{\mathrel{\hbox{\rlap{\hbox{\lower4pt\hbox{$\sim$}}}\hbox{$<$
}}}}

\def\cc{\mbox{cm$^{-3}$}}
\def\tauv{\mbox{$\tau_V$}}
\def\av{\mbox{$A_V$}}
\def\ra{\mbox{$\rightarrow$}}
\def\nhtwo{\mbox{n$_{H_{2}}$}}

\def\HI{H{\smc I}}
\def\HII{H{\smc II}}
\def\m17{M~17}               
\def\cepa{Cepheus~A}               
\def\Htwo{H$_2$}               
\def\HtwoO{H$_2$O}             
\def\orthoHtwoO{o--H$_2$O}             
\def\HtwoeiO{H$_2^{18}$O}             
\def\HtwoCO{H$_2$CO}           
\def\HtwoCS{H$_2$CS}           
\def\Hthreep{H$_3^+$}          
\def\HtwoDp{H$_2$D$^+$}        
\def\DtwoHp{D$_2$H$^+$}        
\def\HthreeOp{H$_3$O$^+$}          
\def\Dthreep{D$_3^+$}          
\def\HCOp{HCO$^+$}             
\def\DCOp{DCO$^+$}             
\def\HthCOp{H$^{13}$CO$^+$}    
\def\HtwCsiOp{H$^{12}$C$^{16}$O$^+$} 
\def\HCSp{HCS$^+$}             
\def\HthCN{H$^{13}$CN}         
\def\HCfiN{HC$^{15}$N}         
\def\HtwCfoN{H$^{12}$C$^{14}$N}  
\def\HNthC{HN$^{13}$C}         
\def\HfoNtwC{H$^{14}$N$^{12}$C}  
\def\HCthreeN{HC$_3$N}         
\def\twCO{$^{12}$CO}           
\def\thCO{$^{13}$CO}           
\def\CseO{C$^{17}$O}           
\def\CeiO{C$^{18}$O}           
\def\twCsiO{$^{12}$C$^{16}$O}  
\def\thCsiO{$^{13}$C$^{16}$O}  
\def\twCeiO{$^{12}$C$^{18}$O}  
\def\thCeiO{$^{13}$C$^{18}$O}  
\def\CtfS{C$^{34}$S}           
\def\thCS{$^{13}$CS}           
\def\twCttS{$^{12}$C$^{32}$S}  
\def\tfSO{$^{34}$SO}           
\def\ttSsiO{$^{32}$S$^{16}$O}  
\def\SOtwo{SO$_2$}             
\def\tfSOtwo{$^{34}$SO$_2$}    
\def\SiO{SiO}             
\def\Ntwo{N$_2$}               
\def\Otwo{O$_2$}               
\def\NtwoHp{N$_2$H$^+$}        
\def\NtwoDp{N$_2$D$^+$}        
\def\NHthree{NH$_{3}$}         
\def\CHthreeCCH{CH$_3$C$_{2}$H}     
\def\CHthreeCN{CH$_3$CN}       
\def\CHthreeOH{CH$_3$OH}       
\def\CHfour{CH$_4$}       
\def\COtwo{CO$_2$}       
\def\thCHthreeOH{$^{13}$CH$_3$OH}       
\def\twCHthsiOH{$^{12}$CH$_3$$^{16}$OH} 
\def\CtwoH{C$_2$H}             
\def\CtwoS{C$_2$S}             
\def\CHp{CH$^{+}$}             
\def\Cp{C$^+$}             
\def\Hp{H$^+$}             
\def\Hep{He$^+$}             
\def\CthreeHtwo{C$_3$H$_2$}    
\def\Jthoh{$J = 3/2 \to 1/2$}
\def\Johoh{$J = 1/2 \to 1/2$}
\def\Jtwel{$J = 12 \to 11$}
\def\Jelt{$J = 11 \to 10$}
\def\Jtn{$J = 10 \to 9$}
\def\Jne{$J = 9 \to 8$}
\def\Jes{$J = 8 \to 7$}
\def\Jss{$J = 7 \to 6$}
\def\Jsf{$J = 6 \to 5$}
\def\Jff{$J = 5 \to 4$}
\def\Jft{$J = 4 \to 3$}
\def\Jtt{$J = 3 \to 2$}
\def\Jto{$J = 2 \to 1$}
\def\Joz{$J = 1 \to 0$}
\def\WCO{W({\rm CO})}
\def\Wtw{W({\rm ^{12}CO})}
\def\Wth{W({\rm ^{13}CO})}
\def\dv{\Delta v}
\def\dvtw{\Delta v({\rm ^{12}CO})}
\def\dvth{\Delta v({\rm ^{13}CO})}
\def\NCO{N({\rm CO})}
\def\Nth{N({\rm ^{13}CO})}
\def\Ntw{N({\rm ^{12}CO})}
\def\NtwCsiO{N({\rm ^{12}C^{16}O})}
\def\NthCO{N({\rm ^{13}CO})}
\def\NthCsiO{N({\rm ^{13}C^{16}O})}
\def\NtwCeiO{N({\rm ^{12}C^{18}O})}
\def\intCO{\int T_R({\rm CO})dv}
\def\inttwCsiO{\int T_R({\rm ^{12}C^{16}O})dv}
\def\intthCsiO{\int T_R({\rm ^{13}C^{16}O})dv}
\def\inttwCeiO{\int T_R({\rm ^{12}C^{18}O})dv}
\def\NHtwo{N({\rm H_2})}
\def\Wtw{W_{12}}
\def\Wth{W_{13}}
\def\kappanu{\kappa_{\nu}}
\def\phinu{\varphi_{\nu}}
\def\taunu{\tau_{\nu}}
\def\dv{\Delta v}
\def\dvFWHM{\Delta v_{FWHM}}
\def\vLSR{v_{LSR}}
\def\Rsol{R_\odot}
\def\Msol{M_\odot}
\def\MMsol{\ts 10^6\ts M_\odot}
\def\MCO{M_{\rm CO}} 
\def\Mvir{M_{\rm vir}}
\def\TAstar{T^*_A}
\def\TAstartwCO{T^*_A(^{12}{\rm CO})}
\def\TAstarthCO{T^*_A(^{13}{\rm CO})}
\def\TAstarCeiO{T^*_A({\rm C}^{18}{\rm O})}
\def\TRstar{T^*_R}
\def\TexCO{T_{ex}({\rm CO})}
\def\Trms{T_{rms}}
\def\d{^\circ}
\def\h{^{\rm h}}
\def\mi{^{\rm m}}
\def\s{^{\rm s}}
\def\mum{\ts \mu{\rm m}}
\def\mm{\ts {\rm mm}}
\def\cm{\ts {\rm cm}}
\def\percm{\ts {\rm cm}^{-1}}
\def\m{\ts {\rm m}}
\def\kms{\rm{\, km \, s^{-1}}}
\def\K{\ts {\rm K}}
\def\Kkms{\ts {\rm K\ts km\ts s^{-1}}}
\def\kHz{\ts {\rm kHz}}
\def\MHz{\ts {\rm MHz}}
\def\GHz{\ts {\rm GHz}}
\def\pc{\ts {\rm pc}}
\def\kpc{\ts {\rm kpc}}
\def\Mpc{\ts {\rm Mpc}}
\def\cmsq{\ts {\rm cm^2}}
\def\pcsq{\ts {\rm pc^2}}
\def\dsq{\ts {\rm deg^2}}
\def\debye{\ts10^{-18}\ts {\rm esu}\ts {\rm cm}}
\def\swash2o{$1_{10} - 1_{01}$}             

\let\ap=\approx
\let\ts=\thinspace

\def\an{{\em Astronomische Nachrichten }}
\def\sci{{\em Science }}
\def\prl{{\em Phys. Rev. Lett. }}
\def\zfa{{\em Zeitschrift fur Astrophysik }}
\def\ba{{\em Baltic Astronomy }}
\def\rmp{{\em Rev. Mod. Phys. }}
\def\rpp{{\em Rep. Prog. Phys. }} 
\def\pasj{{\em Pub. Astron. Soc. Japan }}

\jname{Annu. Rev. Astron. Astrophys.}
\jyear{2007}
\jvol{45}
\ARinfo{1056-8700/97/0610-00}

\title{COLD DARK CLOUDS:
The Initial Conditions for Star Formation}

\markboth{Bergin \& Tafalla}{Cold Dark Clouds}

\author{Edwin A. Bergin 
\affiliation{Department of Astronomy, University of Michigan, 500 Church St. Ann Arbor, MI, 48109, USA; email: ebergin@umich.edu}
Mario Tafalla
\affiliation{Observatorio Astron\'omico Nacional, Alfonso XII 3, E-28014 Madrid, Spain; email: m.tafalla@oan.es}
}


\begin{abstract}
Cold dark clouds are nearby members of the
densest and coldest phase in the galactic interstellar medium,
and represent the most accessible
sites where stars like our Sun are currently being born.
In this review we discuss recent progress in their
study, including the newly discovered
infrared dark clouds that are likely precursors to stellar clusters.
At large scales, dark clouds present filamentary
mass distributions with motions dominated by supersonic
turbulence. At small, sub-parsec scales, a
population of subsonic starless
cores provides a unique glimpse of the conditions prior to stellar birth.
Recent studies of starless cores reveal a combination
of simple physical properties together with a complex chemical
structure dominated by the freeze-out of molecules onto cold dust grains.
Elucidating this combined structure is both an observational
and theoretical challenge whose solution will bring us closer
to understanding how molecular gas condenses to form stars.
\end{abstract}

\maketitle

\section{INTRODUCTION: HOLES IN THE HEAVENS}

``Hier ist wahrhaftig ein Loch im Himmel!'' or ``Here is truly a hole in the heavens!'',
William Herschel was heard by his sister Caroline to exclaim in 1784 when he pointed his telescope towards the constellation Scorpius and viewed the dark nebulae in Ophiuchus \citep{houghton1942}.   Herschel reported his discovery the following year \citep{herschel1785}, and for the next century astronomers debated whether these dark objects where truly voids, left by the stars as theorized by Herschel, or were perhaps nebulous dark regions observed projected upon a dense and bright stellar background.   Our views began to change when Edward Emerson Barnard, who was motivated by Herschel's discovery, published the first modern and systematic photographic survey of the ``Dark Markings of the Sky'' \citep{barnard_survey}.  Barnard argued that his deep photographs provided increasing evidence that many of these dark areas were ``obscuring bodies nearer to us than the distant stars.'' 

Such regions might remain just obscured curiosities were it not for the clear association with star formation which began to be recognized in the mid-point of the twentieth century.  The first person to do so was Bart J. Bok in 1946 \citep{bok48} when he claimed that the dark nebula, in particular compact and nearly round ones that now bear the name ``Bok Globules", are the sites of stellar birth.   It was the advent of modern infrared and millimeter-wave technology in the latter half of the twentieth century that cemented the  relation between Dark Clouds and the formation of stars and planetary systems.

The discovery of molecules in space further revealed that Dark Clouds are made 
of molecular material with \Htwo\ as the dominant constituent \citep{wei63, wil70}.   
Since this finding, the terms Molecular Cloud and Dark Cloud 
are often used interchangeably,
as they refer to the two main characteristics of the clouds: their 
molecular composition and their opaque optical appearance. 
Molecular clouds are dark not because of their hydrogen molecules, but because
of a population of tiny solids (``dust grains'') 
that absorb the optical starlight and lead to high 
visual extinctions (\av\ $> 1^m$). Such a
dimming of the starlight reduces the heating effects from 
external radiation and results in temperatures a few degrees above the 2.7 K cosmic 
background (T $\sim 10$~K).    

For the purpose of this review we will use the term Cold Dark Cloud to refer to those molecular clouds  that are close enough ($< 500$ pc) to be seen in silhouette against the background galactic starlight.  These clouds (e.g. Taurus, Perseus, Ophiuchus, Lupus, ...) are observed to be forming low mass stars either in isolation or in small compact groups and have gas temperatures $\sim 10-20$~K.  This stands in contrast to diffuse clouds in the Milky Way, which contain molecular material but are not forming stars and are not optically opaque with \av\ $\lesssim 1^m$.
At the other end of the spectrum of galactic star formation lie
the more distant and more massive Giant Molecular Clouds that form rich stellar clusters and contain embedded massive stars that heat the surrounding gas to temperatures $>$ 20~K.   

Over the past decade the development of sensitive continuum and heterodyne detectors at millimeter and sub-millimeter wavelengths has led to clear and important advances in our knowledge of the physics and chemistry of these Cold Dark Clouds.  Particular progress has been made in the study of regions prior to onset of star formation, which is the primary focus of this review.
We will first discuss the gains in our knowledge starting from the larger tens to hundreds of parsec scale of molecular  clouds.   We will follow with a detailed discussion of the physics and chemistry of the much smaller, $\sim$0.1 pc, scale of molecular cloud cores, where recent research has clarified a number of outstanding issues.
We conclude with a discussion of a newly discovered class of cold ($T_{gas} \sim 10-20$~K) dark clouds: infrared dark clouds that represent a new population of dense molecular regions and are believed to be the precursors to stellar clusters and massive stars.   

 Space constraints prohibit this review from complete coverage of the formation of clouds, stars, and planets.  For more details we refer the reader to recent reviews of star formation \citep{lar03} (see also the review chapter by McKee and Ostriker, this volume), the role of turbulence \citep{mac04, elmegreen_scalo}, probes of physical conditions \citep{evans_araa}, gas/dust chemistry and star formation \citep{vdb_araa, ehr00, vd_araa}, and the entire Protostars and Planets V volume \citep{ppv},
 especially the papers by Ceccarelli et al., Di Francesco et al., and Ward-Thompson et al. that contain complementary reviews of this topic and include references that for reasons
of limited space could not be mentioned here. Recent accounts of how the new 
observations with the Spitzer Space Telescope are advancing our
knowledge of star formation in dark clouds can be found in \citet{wer06} 
and \citet{all06}.

While recent progress clearly warrants this review,
the year 2007 presents an auspicious occasion for an overview of dark clouds.  It is 150 years since the birth of E. E. Barnard in Nashville in December 1857, 101 years since the birth of B. J. Bok in the Netherlands, and 80 years since Barnard's Photographic Atlas of Selected Regions of the Milky Way was published posthumously \citep{barnard_atlas}.  
It is also the centennial of Barnard's beautiful exposure of Taurus-Auriga, which is shown in Fig. 1.   The hundred years
that separate us from Barnard's picture of Taurus have brought
us a wealth of knowledge on the physics and chemistry of these
dark patches. As we will see in this review, however, dark clouds
remain in some aspects as mysterious  and fascinating as
when Barnard pointed his camera to them.

\begin{figure}
 \centerline{\psfig{figure=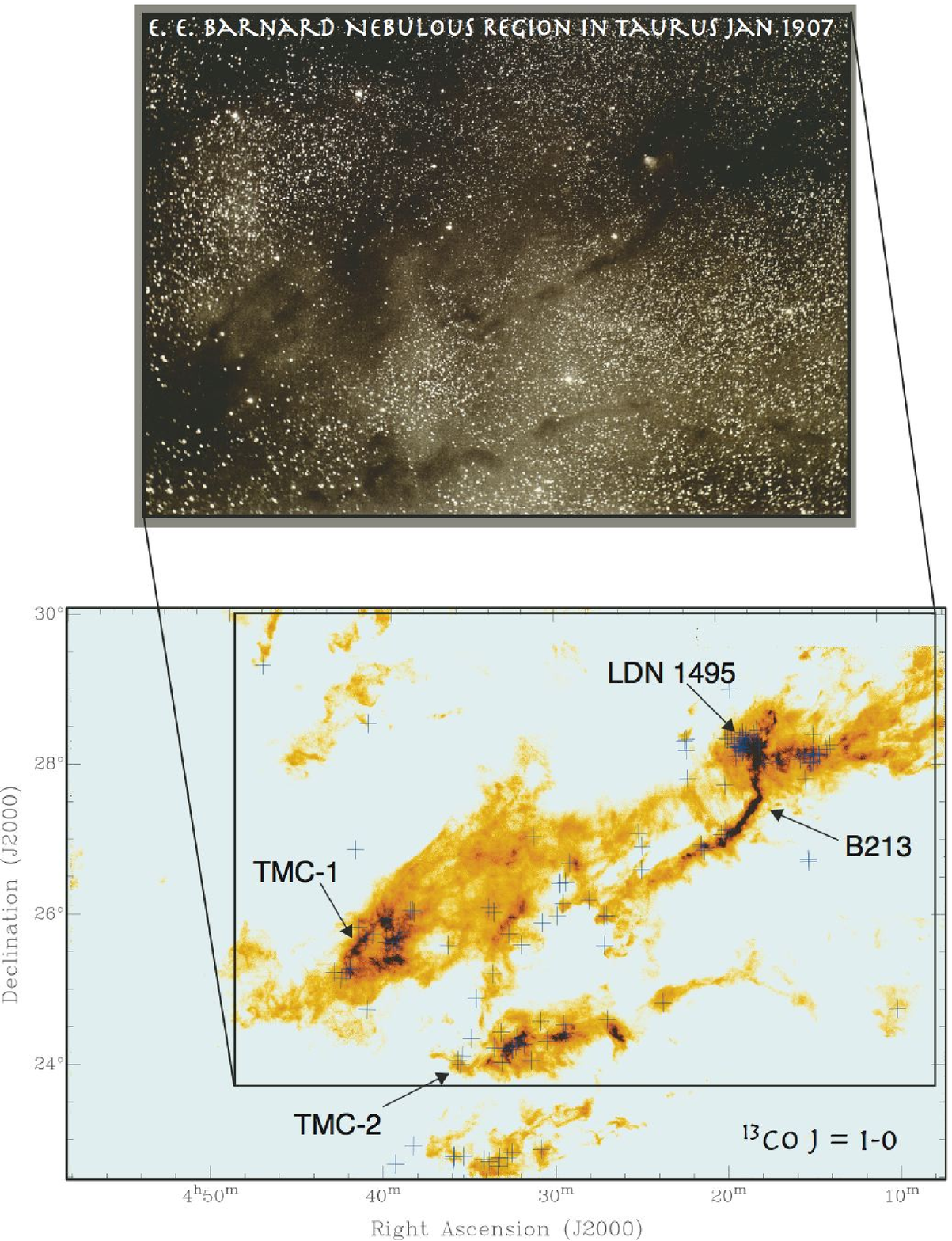,height=6.5in}}
 \caption{
 {\it Top Panel:} Photographic image of the Taurus Molecular Cloud taken by E. E. Barnard \citep{barnard_survey}.  His notes state that, ``{\em very few regions of the sky are so remarkable as this one.  Indeed the photograph is one of the most important of the collection, and bears the strongest proof of the existence of obscuring matter in space}''.   Courtesy of the Observatories of the Carnegie Institution of Washington. {\it Bottom Panel:} \thCO\ J=1--0 integrated emission map of the same region obtained using the Five College Radio Astronomy Observatory.  Crosses mark the location of known protostellar objects and the emission color scale ranges from 0.5 -- 10 K km s$^{-1}$.  Image kindly provided by PF Goldsmith in advance of publication.
 \label{fig:taurus}}
\end{figure}

\section{LARGE SCALE STRUCTURE OF COLD DARK CLOUDS}

\subsection{Recent Studies}

A number of technical developments over the last decade have rapidly increased
our ability to study dark clouds. Large-scale digital images in the
optical, infrared, and radio regimes have become available in a routine
or quasi-routine basis, and their analysis allows studying 
degree-sized portions of the sky at arcminute or better resolution.
In the following sections we review some of these developments
ordered
by wavelength, and we briefly discuss the analysis techniques
used to exploit this new wealth of data. Thanks to this recent progress,
the study of cloud structure is currently enjoying a new golden age.

\subsubsection{OPTICAL AND IR DATA}

The Digitized Sky Survey (DSS) 
and the USNO Precision Measuring Machine project 
\citep[USNO-PMM;][]{mon96} have provided multi-band, 
all-sky images at optical wavelengths
with accurate calibration and registration. Using different techniques,
these data can be processed automatically 
to create large-scale images of dark clouds.
\citet{cam99} has used USNO-PMM data together with a star-counting method
to produce extinction ($A_V$) maps of up to $\sim$ 250 square degrees for 24
dark clouds including Taurus, Perseus, and Ophiuchus. He has used 
an automatic adaptive grid with a fixed number of stars per cell to
avoid empty pixels and to achieve a maximum extinction 
of around $A_V \sim 7$. This adaptive cell method, however, degrades the
angular resolution of the images from about 1 arcminute
in the outer parts of the clouds to about 10 arcminutes in the most
opaque areas. A similar star-counting technique, but with a fixed cell spacing, 
has been used by \citet{dob05} to make an atlas of dark clouds using DSS-I data.
These authors identify 2448
dark clouds toward the galactic plane ($b \leq 40^\circ$) and present
maps with angular resolutions of $6'$ and $18'$ that are publicly available
(\url{http://darkclouds.term.jp}).

Near infrared observations can be used to extend
the optical extinction measurements to the most opaque regions of 
clouds thanks to the wavelength dependence of the dust absorption.
The recent development of large-format infrared arrays has
made it possible to apply this technique to increasingly large regions 
of the sky, and thus complement the work carried out at
optical wavelengths.
In addition to applying the star-count method, which provides
an estimate of the average extinction inside a grid cell,
multi-band NIR observations can be used to provide a direct determination of
the extinction towards individual
background stars. This determination uses the small range of
NIR colors spanned by the background stars (mostly giants), which
for practical purposes and with little error can be
assumed constant. The Near Infrared Color Excess (NICE) method
of \citet{lad94} uses H and K observations of background
stars together with an estimate of their intrinsic (H-K) color
from a nearby unobscured control field to derive $A_V$ for each observed star.
As an application of this method, \citet{lad94} produced an accurate map
of the extinction towards the IC~5146 dark cloud
with a resolution of 1.5 arcminute (also \citealt{lad99}).
An improvement of this technique is the
NICER (NICE Revisited) method of \citet{lom01}, which uses
observations in additional NIR bands to reduce the variance of the $A_V$
estimate.
By applying this method to the Orion data of the all-sky 2MASS
survey \citep{skr06}, these authors show that it is possible to derive large-scale
cloud structure with $5'$ resolution to a limiting extinction of $A_V\approx 0.5$
(better or comparable to optical star-count work and below
the threshold for molecular CO formation). Future use of this technique
with 8m class telescopes is expected to provide accurate extinction maps with
two orders of magnitude dynamical range and a
resolution better than $10''$ \citep{lom01}.

\subsubsection{RADIO DATA}

In parallel with the technical advances in NIR cameras, millimeter and 
submillimeter-wavelength bolometer arrays have grown in size and sensitivity 
over the last decade. Arrays like SCUBA on the JCMT, MAMBO 
on the IRAM 30m telescope, and Bolocam on the CSO
have made it possible to map systematically
the thermal emission of cold dust from dark clouds \citep[e.g.][]{mot98,joh00,eno06}.
In contrast with their NIR counterpart, however, millimeter (mm) and 
submillimeter (submm) observations 
of dark clouds are sensitive to a narrower range of dust temperatures
and suffer from severe instrumental
limitations. The need to chop the telescope between sky and reference
positions filters out part of the large-scale emission, which is usually
weak enough to be close to the detection limit of the array. This means 
that bolometer observations miss most of the extended emission from 
a cloud while they recover its full small-scale structure. The 
future 
SCUBA-2 array at the JCMT, with its expected thousand-fold increase in 
mapping speed and high sensitivity for extended emission,
promises to revolutionize the field of dust continuum mapping of
dark clouds.

Molecular-line observations, on the other hand,
provide information about the gas component of clouds,
in particular about their velocity structure, density, temperature, and chemical
composition. These data are complementary to the dust extinction 
and emission measurements, and they are required to understand, among other
issues, the equilibrium status and the chemical evolution of dark clouds. Progress
in mm-line receiver technology over the last ten years has also improved our
ability to map large areas of the sky with increasing angular and spectral
detail. Low resolution ($\approx 10'$) CO imaging of the
entire Milky Way has been carried out with the CfA 1.2m telescope \citep{dam01},
and systematic mapping of selected clouds at higher 
resolution ($\approx 2'$) has been performed in CO isotopologues
with the NANTEN/Nagoya 4m telescope \citep[e.g.,][]{tac02}
and the KOSMA 3m telescope \citep[e.g.,][]{sun06}. The 32-beam
SEQUOIA array (and its predecessor, the 15-beam QUARRY array), 
operating at 3mm on the FCRAO 14m telescope, have been used to make
megapixel images
with sub-arcminute resolution of nearby clouds like Taurus 
\citep[Fig. 1 and][]{gol05b} and further out regions like the Galactic Ring
\citep{jac06} and the outer Galaxy \citep{hey98}. 
This array has also provided line data for the
Coordinated Molecular Probe Line Extinction and Thermal Emission
(COMPLETE) project \citep{rid06},
which has the goal of combining the different IR and radio techniques of
dark cloud mapping to study a series of dark
clouds common to the Spitzer Space Telescope Legacy Program
``From Molecular Cores to Planet Forming Disks'' 
\citep[c2d,][]{eva03}. The potential of heterodyne arrays, 
however, has not been fully exploited yet because the
reduced number of pixels in the receivers limits the 
large scale observations to the bright lines of CO isotopologues. 
Further progress in this 
technology is still crucial to our understanding of
the large-scale physics and chemistry of dark clouds.

\subsection{Observational Properties of Dark Clouds}

The dark cloud data 
obtained using the new instrumentation
complements a large body of previous work
summarized in previous reviews like those by \citet{gol87},
\citet{lad93}, and \citet{mye95}.
These new data have helped greatly to refine
our understanding of the dark cloud properties,
although, as we will see, a number of important
uncertainties still remain.
In the following sections, we review some of the basic
properties of clouds with emphasis on 
recently clarified issues and general characteristics
from an observational point of view, concluding with a 
brief review of the issues whose solution
remains pending. Because 
of space limitations, we cannot do justice to all the
excellent studies of individual clouds carried out
with the optical, IR, and radio instrumentation discussed 
before.

\subsection{Shapes}

A simple inspection of dark cloud images obtained by any of the 
methods discussed previously reveals that clouds come in a variety of
sizes and shapes. In general, dark clouds have highly
irregular edges, and their overall appearance is filamentary and often
wind-blown. The presence of long, well-defined filaments 
was emphasized already a century ago by \citet{bar07} when 
he noted, describing the Taurus plate of Fig. 1, 
that ``among the most surprising 
things in connection 
with these nebula-filled holes are the vacant lanes that so
frequently run from them for great distances." 
Indeed, some of the filaments in Taurus can be followed 
for more than 4 degrees or 10~pc,
and similarly thin and long structures can be seen in many other
clouds like Ophiuchus, Lupus, and Orion
both in optical plates \citep[e.g.,][]{sch79}
and radio images \citep[e.g.,][]{bal87,joh99}.

In many clouds including Taurus and Ophiuchus, the
length of some filaments is comparable to
the full extent of the cloud. Typically, a
cloud contains two or three long filaments
that are either parallel or converge at a low
angle in a massive condensation
that often contains
an active cluster-forming site \citep{tac02,bur04}.
The velocity field of some filaments,
in addition, seems rather coherent \citep[see][
for a study of Ophiuchus]{lor89}. This combination of spatial length
and velocity coherence for some of the filaments
suggests that their presence is intrinsic to 
the cloud structure and not the result of later evolutionary factors
like star-formation activity, which would produce a more
chaotic, small-scale mass distribution.
Dark clouds, therefore, seem to be born with a
filamentary distribution of material that extends 
over a number of parsecs. 
As clouds evolve and form stars, the products of star
formation inherit the filamentary distribution
of the parental gas \citep{har02}.

\subsection{Mass Distribution}

Maps of dark clouds provide more than just information on shapes.
Measurements of dust extinction or gas column density at each
cloud position can be used to estimate the amount of material 
under different physical conditions, and in particular
at different densities.
In his extinction study of 24 dark clouds,
\citet{cam99} finds that most of them present a
similar power-law relation between the mass
enclosed in an iso-extinction contour and the extinction,
although the $A_V$ range of this optical study is restricted
to $A_V \approx 1$-5.
Despite this limitation, the measurements 
illustrate how most of the material in a cloud lies at relatively low
extinctions, and it must therefore reside
in the form of low density gas. Using more sensitive
NIR extinction data, \citet{alv99} 
find that only 25-30\% of the total mass in
the IC~5146 and L977 cloud lies at
$A_V > 10$, while \citet{cam02} finds
that this percentage is slightly more than 10\% 
for the North America Nebula. A much smaller 
value ($\sim 1$\%) has been recently reported for the Pipe 
Nebula, also from IR extinction measurements 
(\citealt{lom06}, also \citealt{lad06}).
Similar low fractions of dense material are 
indicated by large-scale maps of dust emission
in clouds actively forming stars.
\citet{joh04} and \citet{you06} find
that only between 1 and 2\% of the mass in the Ophiuchus cloud
is associated with dense, submm continuum emission,
and that this emission is only detected towards regions of 
extinction larger than 10-15 mag. This is illustrated in Figure
2, which shows the IR-extinction and submm-emission maps
of Ophiuchus from \citet{you06}.

\begin{figure}
 \centerline{\psfig{figure=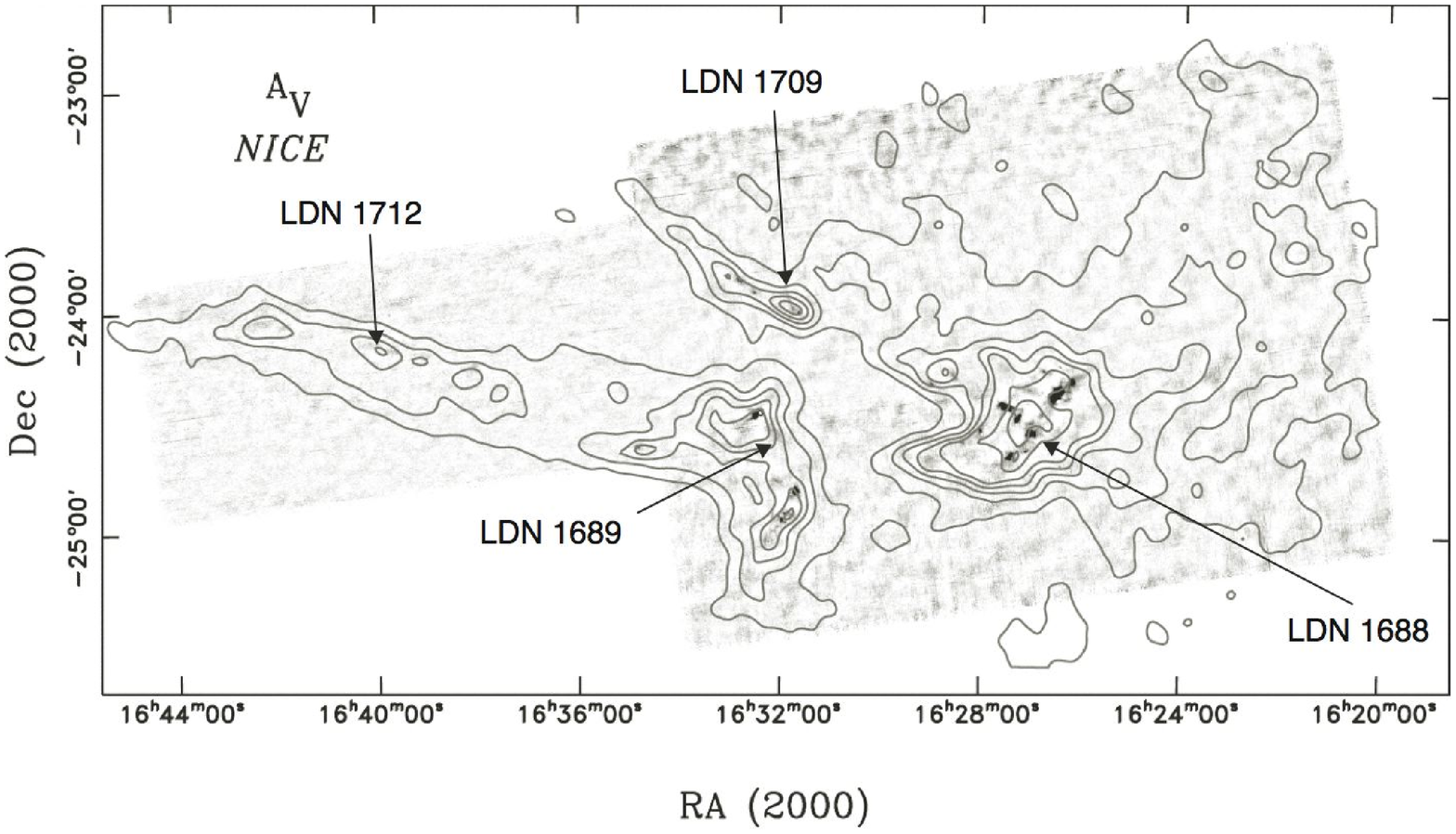,height=3in}}
 \caption{Maps of IR extinction (contours) and submm emission (greyscale) for
 the $\rho$ Ophiuchus dark cloud illustrating how the dense gas, responsible for
 the submm continuum emission, represents a small fraction of the total amount (and an 
 even smaller fraction of the total volume). Most of the gas in the cloud is in
 the form of low density, low extinction gas. Contours are
 at $A_V = 2, 4, 6, 8, 10, 15$, and 20 mag with an effective resolution of
 $5'$. Figure from \citet{you06}. 
\label{fig:oph}}
\end{figure}

A related measure of how the mass of a cloud is distributed
among different regimes is the statistics of extinction values.
\citet{rid06} find lognormal
distributions for the number of pixels as a function
of extinction in both Perseus and Ophiuchus, a 
distribution that is in agreement with the expectation
from turbulent models of cloud formation \citep{vaz94,ost01}.
\citet{lom06}, in their
study of the Pipe Nebula, find that while a lognormal distribution
provides an approximate general fit to the data, the observed histogram
has significantly more structure that can only be reproduced
with additional components.

\subsection{Velocity Structure}
\label{sec:cloud_veloc}

The presence of supersonic motions in dark clouds has been
known since the first molecular-line studies 
revealed non-gaussian line shapes and velocity differences across
clouds of the order of 
1 km s$^{-1}$ or more (sound speed is 0.2 km s$^{-1}$ for 
10 K gas). These motions are inferred to be 
turbulent from the lack of systematic patterns
like rotation, expansion, or infall
\citep[e.g.,][]{hei76}, and from
the existence of a systematic power-law relation between the 
velocity dispersion inside a cloud $\sigma$ and its physical size
$L$ ($\sigma$(km s$^{-1}$)= 1.10 $L$(pc)$^{0.38}$, \citealt{lar81}),
which is reminiscent of the classical Kolmogorov law
of (incompressible) turbulence. 
Understanding and characterizing these turbulent
motions, which affect not just dark clouds but the whole
interstellar medium at almost every scale, is 
a formidable task. In contrast with their
laboratory counterparts, turbulent motions in the interstellar 
medium are supersonic, compressible, and likely of
magnetohydrodynamical nature. A rich field of both 
analytic and numerical work has been developed
to tackle this problem, and a review of its results
exceeds the scope of this paper. 
We refer the reader to the excellent recent review by 
\citet{elmegreen_scalo} for an in-depth view of interstellar
turbulence, especially from
a theoretical point of view, and for an extensive list of references. 
Here we will concentrate on the main observational results of 
recent studies of dark cloud kinematics.

When studying the kinematics of cloud gas, it is necessary to
take into account its dependence on 
density. Models of supersonic turbulence show that 
the highest density gas moves slower than the 
low density material, as it arises from the convergence
of flows \citep[e.g.,][]{pad01}, and this
velocity segregation seems to be present in clouds.
The densest regime in a cloud corresponds to the
dense cores, which are known to have 
very low levels of turbulence and subsonic internal motions 
\citep[e.g.,][]{myers83}. Dense cores seem therefore to lie 
at the bottom of the hierarchy of cloud kinematics, 
and to represent structures dominated by the dissipation
of turbulence \citep[e.g.][]{goo98}. In \S~\ref{sec:physics}
we will present a more complete summary of 
the physical properties of dense cores, and defer 
the discussion of dense gas kinematics to that section.

The kinematics of the low density gas in dark clouds
at the small scales typical of dense cores ($\sim 0.1$ pc)
has been studied in detail by \citet{fal98}.
These authors selected 3 regions in nearby clouds 
and mapped them in different CO isotopologues
over several 
arcminutes (tenths of a parsec for the clouds distance)
with the IRAM 30m telescope. 
Each selected region contains a starless dense core, although 
the CO emission is
dominated by the more diffuse material because of optical depth
and depletion effects (see \S~\ref{sec:obschem}). Even at their limit of 
angular resolution (about 1700 AU), \citet{fal98} find 
unresolved velocity structure, especially at the highest speeds.
This fastest gas appears as prominent wings in non-Gaussian CO spectra 
with velocities of about 1 km s$^{-1}$
and has an extremely filamentary distribution with
aspect ratios larger than about 5. In contrast with the 
dense ($10^5$ cm$^{-3}$), subsonic, gas in a core,
the fast CO-emitting gas (with a density of a few  $10^3$ cm$^{-3}$)
presents large shears and velocity dispersions at the smallest 
resolvable spatial scales. The small regions of high shear 
may represent locations of enhanced dissipation of
turbulence \citep{pet03}.

The large-scale distribution of turbulence in clouds has been
studied by a number of authors using different techniques.
In addition to the classical linewidth-size relation of 
\citet{lar81}, methods used to analyze cloud
kinematics include the statistics of the 
centroid velocity fluctuations \citep{mie99}, the 
spectral correlation function \citep[SCF,][]{ros99}, the
principal component analysis \citep[PCA,][]{hey97}, and
the velocity channel analysis \citep[VCA,][]{laz00}.
A common goal of these methods is to identify and 
extract from maps of molecular-line
emission a number of statistical properties 
of the velocity field
that can be compared with the
expectations from analytic theory (like Kolmogorov's 
law) or from numerical simulations of hydrodynamic and
magnetohydrodynamic turbulence 
\citep[e.g.,][]{oss02,bal06}.
Two main issues addressed by these
studies are the way in which the turbulent energy is distributed between 
the different spatial scales 
(as described by the energy spectrum $E(k)$)
and the nature of the driving agent 
\citep[for more details, see][]{elmegreen_scalo}.

A recent 
example of the above type of work is the principal component
analysis of 23 fields of the FCRAO CO Outer Galaxy Survey
by \citet{bru02}.  Using these and additional data, \citet{hey04} 
have found
an almost universal power-law form for the energy spectrum 
of clouds, with very little cloud-to-cloud scatter 
(10-20\%) not only in the power law index but in the normalization coefficient.
Such little scatter would help explain the emergence of Larson's law when
combining data from an ensemble of clouds \citep{lar81}. 
By combining numerical simulations of supersonic turbulence
with an analysis of the velocity line centroid, the linewidth-size relation,
and the PCA determinations of turbulence properties,
\citet{bru04} suggest that observations of line emission 
from clouds are best fitted by compressible,
shock-dominated (Burgers) turbulence. They also suggest that
either clouds are recently formed or they are continually driven on large
scales.
These suggestive results still need confirmation from the systematic
observation of a large number of nearby clouds and from the
realistic modeling of the combined
effect of cloud kinematics, chemistry, and radiative transfer.
The extreme complexity of turbulent motions in clouds, the diverse
mechanisms likely involved in their driving and dissipation,
and the challenge of obtaining high-quality molecular-line data
of whole clouds will surely keep challenging theorists and observers
alike for a number of years.

\subsection{Internal Structure}
\label{sec:cloud_struct}

A common feature of dark clouds is that they show 
a partly hierarchical structure, with smaller
subunits appearing within large ones when 
observed with increasing spatial resolution. 
To characterize this structure,
two different approaches have been generally followed,
depending on whether the discrete or the continuous nature 
of the structure is emphasized. 
In the first approach, the cloud is assumed to be composed of
subunits, referred as clumps and defined as
coherent regions
in position-velocity space that may contain significant
substructure \citep{wil00} (see below and Table 1 for typical
clump parameters).
These clumps are identified from the data using an 
automatic clump-finding algorithm that simplifies
the process and avoids human bias.
The most popular algorithms in use are
GAUSSCLUMPS, developed by \citet{stu90},
and CLUMPFIND from  \citet{wil94}. These
algorithms follow different approaches to 
identify and characterize clumps, but they generally agree
in the derived properties, especially in the intermediate 
and high mass end \citep{wil94}. 
When they are applied, usually to CO isotopologue data,
the spectrum of clump properties is found to follow a 
power law in the range for which the data are complete. The mass of
the clumps, for example, commonly presents
a distribution of the form $dN/dM \sim M^{-\alpha}$, where
$\alpha$ lies in a narrow range between 1.4 and 1.8
for both dark clouds and GMCs \citep{bli93,kra98}. As a reference,
and in this form, we note that the initial mass function 
of stars follows a steeper power law with a slope 
of $\alpha=2.35$ \citep{sal55}. The clump distribution, therefore,
contains most of the mass in massive clumps, while the stellar distribution
has most of the mass in low-mass objects.

The second approach to characterize cloud structure 
assumes that the cloud is
self-similar, at least over some range of scales, 
and applies the concepts of fractal geometry. 
In this geometry, the most characteristic parameter 
of an object is its fractal (or Hausdorff) dimension, which 
is in general a non integer number \citep{man83}. 
The boundary of a cloud, for example, has 
a fractal dimension $D_p$ that
can be estimated from the relation between the enclosed area ($A$) and 
the perimeter length ($P$): $P\sim A^{D_p/2}$ \citep{lov82}. 
For simple curves in a plane, like circles or ellipses,
this area-perimeter relation 
gives the expected value of $D_p=1$ for the dimension 
of a line. For highly fragmented or convoluted
curves, the expression gives a value that is larger than 1
and that approaches 2 as
the curve fills a larger fraction of the plane. 
\citet{bee87} used the
above area-perimeter relation for the optical boundary of
24 dark clouds and derived an average boundary dimension
of 1.4, which suggests that clouds are fractal. 
Using a similar method for the iso-contours of
a series of CO maps, \citet{fal91} have 
derived a bounday dimension $D_p = 1.36\pm 0.02$ for the Taurus molecular
cloud, almost the same value measured by 
\citet{sca90} from IRAS maps.
Similar values for the fractal dimension have been obtained
for the more diffuse cirrus clouds \citep[$D_p = 1.3$,][]{baz88} 
and even for the atmospheric clouds on Earth \citep[$D_p=1.35$,][]{lov82}.
As these dimension estimates use plane-of-the-sky views
or maps of the objects, they only provide the fractal dimension
of the clouds projection. To infer the intrinsic (volume)
fractal dimension, the area-perimeter value is generally
increased by 1 \citep{bee92}, although this procedure may
only provide a lower limit to the true dimension \citep{san05}.
With this assumption, the area-perimeter determinations imply a
typical cloud fractal dimension of around $D=2.4$.

The very different approaches of the clump-finding and
fractal analysis would seem to suggest that the two methods
provide contradictory views of the cloud internal structure. 
This is however not the case, as the power-law
behavior found in the spectra of clump properties
already suggests a scale-free structure consistent with 
a fractal geometry (\citealt{elm96}, also \citealt{stu98}).
Indeed, \citet{elm96} show that
it is possible to derive clump mass distribution functions
as those found by the GAUSSCLUMPS and CLUMPFIND algorithms
from fractal cloud models of (volume) dimension $D = 2.3 \pm 0.3$.
In this view, the clump-finding algorithms identify 
the peaks of the fractal intensity distribution of a cloud.
Due to the threshold imposed in the density or
column density by the observations, a power-law spectrum of 
cloud sizes results, and this produces a power-law distribution
of clump masses even if the gas density follows a
lognormal distribution \citep{elm02}.

The self-similar behavior revealed by both the power-law
clump decomposition and the fractal description seems to apply 
inside a given range of spatial scales. This range
most likely represents the regime dominated by
turbulent motions, as turbulence produces self-similar
structures in a natural way \citep[e.g.,][]{elmegreen_scalo}.
At large scales, the presence of well-defined features like
filaments or shells indicates a deviation from self similarity 
that is likely caused by the mechanisms responsible for the cloud
formation \citep[e.g.,][]{elm96}.

 At small scales,
in the gravity-dominated regime, the self-similar picture is also 
expected to break down.
No break point from self similarity has yet
been found in the lower density gas, and \citet{fal98}
report unresolved turbulent structures at
scales as small as 0.02~pc from their study of the highest
velocity gas in the Taurus molecular cloud \citep[also][]{pet03}.
For the dense star-forming gas, on the other hand, \citet{lar95}
has inferred a break down of the self-similar behavior in the form of a 
discontinuity in the slope of the surface density
of companions for young stars in Taurus.
This break point divides the binary regime from the clustering regime, 
and it occurs approximately at a radius of 0.04~pc, which 
coincides with the Jeans length of the cloud. Such coincidence suggests 
that for scales smaller than this length, 
thermal pressure provides the dominant support against 
gravity, while for larger scales, turbulent and magnetic pressures
are important \citep{lar95}.
Followup work by \citet{sim97} shows similar
break points in the density of stellar companions for Ophiuchus
and the Orion Trapezium, although the scale-size of the break point
does not follow the expected Jeans mass. A larger value for
the break point in the self-similar behavior of Taurus has been
proposed by \citet{bli97} based on $^{13}$CO observations, although
optical depth 
effects and molecular freeze out (section \S~\ref{sec:obschem})
may have affected this estimate. 

There is further evidence for a change in the structure of dark clouds
at small scales. As mentioned before, the
clump mass spectrum found from
large-scale CO observations is relatively flat, and follows a 
power law of the form $dN/dM \sim M^{-\alpha}$ with 
$\alpha = $ 1.4-1.8. The mass spectrum of 
millimeter/submillimeter dust continuum peaks, on the other hand,
presents a steeper slope of $\alpha=$2.0-2.5 for masses 
larger than about 1 $M_\odot$ \citep{mot98,tes98,joh00,joh01}.
This spectrum, which at least for
the case of Ophiuchus flattens to about $\alpha = 1.5 $ for 
lower masses \citep{mot98,joh00},
mimics the shape of the initial mass function (IMF) of stars 
\citep{sal55,mil79}, in contrast with the clump mass spectrum. 
Such similarity to the IMF 
suggests that stellar masses may be determined by  
the same process that fragments the molecular gas at the
smallest scales. 

The above discontinuity in the slope $\alpha$
still needs confirmation
with a single observational technique, as each
side of the break point has been observed with
a different tracer (CO for large scales and mm/submm 
continuum for small scales). If confirmed, the
discontinuity will suggest a
change in the physics of the cloud gas at 
the tenth of a parsec scale, and this
can provide the basis for an empirical description
of the different levels of cloud structure.
\citet{wil00} have recently proposed one such
a categorization of structure in terms of clouds,
clumps, and cores, with clumps being defined (as before) by their
velocity coherence 
and cores being defined as gravitationally
bound, single peaked regions out of which individual stars
(or simple stellar systems) form. Alternative definitions
have already been proposed \citep[e.g.,][]{gol87,mye95}, and 
it is also possible that any description
of cloud structure using only a few elements 
is too simple to capture the continuous properties of
cloud gas \citep{rod05}.  Still, the need to name and characterize
the different levels in the hierarchical structure
of clouds makes it necessary to have a well accepted terminology, 
and the categorization by \citet{wil00} matches
the spirit of current usage. Thus, using this convention, 
we present in Table~\ref{tab:cores_clumps} a summary of the
main physical parameters of clouds, clumps, and cores.
We stress that these units are still loosely defined and that
their properties may be sensitive to the tracer used in the
measurement and can vary with cloud. Future improvements in our 
understanding of cloud chemistry and structure will help further refine 
the above description.

\begin{table}
\def~{\hphantom{0}}
\caption{Properties of Dark Clouds, Clumps, and Cores}\label{tab:cores_clumps}
\begin{tabular}{@{}lccc@{}}
\toprule
& Clouds$^a$ & Clumps$^b$ & Cores$^c$\\
\colrule
Mass (M$_\odot$) 		& $10^3$--$10^4$& 50--500  	& 0.5--5\\
Size (pc)       		& 2--15  	& 0.3--3	& 0.03--0.2\\
Mean density (cm$^{-3}$)	& 50--500	& $10^3$--$10^4$& $10^4$--$10^5$\\
Velocity extent (km s$^{-1}$)	& 2--5		& 0.3--3	& 0.1--0.3\\
Crossing time (Myr)		& 2--4		& $\approx 1$	& 0.5--1\\
Gas temperature (K)		& $\approx 10$	& 10--20	& 8--12\\
Examples			& Taurus, Oph,	& B213, L1709	& L1544, L1498,\\
				& Musca		& 		& B68\\
\botrule
\end{tabular}
\\
$^a$ Cloud masses and sizes from the extinction maps by \citet{cam99}, velocities and
temperatures from individual cloud CO studies \\
$^b$ Clump properties from \cite{lor89} ($^{13}$CO data) and \cite{wil94} (CO data) \\
$^c$ Core properties from \citet{jij99}, \citet{cas02}, \citet{mot98}, and individual 
studies using NH$_3$ and N$_2$H$^+$ \\
\end{table}

\subsection{Magnetic Field}
\label{sec:cloud_bfield}

The magnetic field of a cloud is probably its most difficult property to
measure. The line of sight strength of the field can only be directly determined
observing the Zeeman splitting of line transitions, and its plane of the sky
direction can be estimated via polarization measurements of background
stars, dust emission, or spectral lines. All these measurements
require observations that combine stability and high signal 
to noise, and are therefore difficult and time consuming 
(see \citealt{hei93}
for an in depth review of observational techniques of magnetic field 
measurements). Although the magnetic field only acts directly on charged
particles (electrons, ions, and charged dust grains), its presence
can be felt by the neutral material through collisions. Under most
dark cloud conditions, this ion-neutral coupling is
highly efficient and, except for the densest regions, the
field is expected to be frozen to the gas
(see \citealt{mck93} for a review of the basic 
magnetic field theory). Because of this, the 
ability of the magnetic field to counteract the action of
self-gravity can be critical to the equilibrium balance
of dark clouds.

The most straightforward way to map the large-scale orientation of
the magnetic field in a cloud is to measure the polarization
of light from background stars, which results from the dichroic extinction 
by aligned, elongated dust grains \citep[e.g.,][]{dav51}.
Large-scale maps of polarization for the Taurus cloud have been 
presented by  \citet{mon84} and \citet{hey87}. \citet{goo90}
have produced maps of the optical polarization
for Perseus, Taurus, and Ophiuchus using a combination of
their own measurements with previous data. From these maps, they find 
that in Taurus and Ophiuchus the pattern of polarization is highly 
regular over scales of about 10 pc, while
the Perseus measurements seem to suffer from the superposition of two 
components along the line of sight. In Taurus and Ophiuchus, the large
scale filaments are neither completely parallel nor completely 
perpendicular to the global
polarization pattern, suggesting that the magnetic field does not
dominate the cloud structure on large scales \citep{goo90}.
Optical polarization data from other clouds offer a mixed picture.
Clouds generally present fields with a well-defined mean direction
but significant dispersion \citep{mye91}, and cases
where the field is parallel or perpendicular to a filament exist
(e.g., the long filamentary Musca cloud presents an ordered field 
perpendicular to the long axis of the filament, see \citealt{per04}).

While optical polarization measurements are necessarily limited
to the low-extinction parts of a cloud, so the background stars
are still bright enough to have their polarization measured,
IR polarization data could in principle sample 
more opaque regions and therefore provide a deeper
view of the inner magnetic field. Unfortunately, this seems not to be
always the case. For the Taurus cloud, for example, there is evidence that 
the dust loses its polarizing power at depths larger than
about $A_V=1.3$, so IR polarization measurements offer little
improvement over optical data in this region \citep{goo95,arc98}.
This loss of polarizing power probably results from 
a combination of a change in the alignment
properties of grains and a change in their optical properties because
of coagulation and mantle growth, although the exact mechanism is not
yet well understood \citep{laz97,whittet01}.

Better sampling of the innermost cloud regions is expected from
polarization measurements of the millimeter/submillimeter emission
from dust, although they also suffer often from drops in
polarizing power towards the densest areas \citep{mat01}.
Because of limitations due to sensitivity and spatial filtering,
large scale studies of the dust polarization
have concentrated on the brightest objects
like the Orion filament  \citep{mat00,hou04}
instead of the more nearby and colder dark clouds (see \S~\ref{sec:cores_bfield}
for submm-polarization studies of dense cores). As in the case of the 
optical/IR polarization measurements, the limited submm measurements
do not show a very strong correlation between the magnetic field direction 
and the mass distribution in the clouds \citep[e.g.,][]{hou04}.

Zeeman effect observations of dark clouds provide an estimate of the strength
of the (line of sight component of the) magnetic field. The high
signal-to-noise required by these measurements and the low
spatial resolution achieved, because of the low frequency of the 
transitions, means that no large-scale maps of Zeeman effect
measurements are currently available. Observations are commonly restricted
to single pointings, which offer only a limited sampling 
of the magnetic field strength in clouds. In addition, and due to the 
need for strong emission, the pointings tend to be selected 
towards the densest regions. Still, coming up with meaningful detections
is challenging, and the OH Zeeman observations of dark clouds by \citet{cru93}
produced one clear detection out of 12 positions observed (mostly
in Taurus and Ophiuchus).  These data, together with the observations
of \citet{tro96} suggest typical magnetic field strengths
of order of 10 $\mu$G or less. 

A main issue for the studies of the magnetic field strength in cloud gas
is the importance of the field in the dynamics and balance of the cloud.
A convenient parameter to characterize this importance is the 
observed ratio between the mass and the magnetic 
flux normalized to the critical value $\alpha G^{-1/2}$, where $G$ is the gravitational constant and $\alpha \sim 0.13$ (see \citealt{mck93}
for a full discussion and equations).  If the observed mass-to-flux
ratio exceeds the critical value, the cloud is said to be supercritical
and the magnetic field cannot prevent collapse. If the ratio
is lower than the critical value, the cloud is called subcritical 
and gravity is stabilized by the magnetic field.
A detailed compilation and analysis of the currently available 
data on magnetic field strength in molecular clouds has been
presented by \citet{cru99}, and an update of this work
appears in the review by \citet{hei05}. 
According to this analysis, the magnetic field strength is 
just at the level of being critical, i.e., at the boundary between being
dynamically important and not. Even in this best-to-date analysis,
there are important caveats due to the large number of non-detections
and the need to correct geometrically the mass-to-flux ratio
(multiplying by 2 in case of a sphere or by 3 in case
of a sheet, see \citealt{bou01} and \citealt{hei05}). The values, in 
addition, are likely more representative of the densest regions
than of the extended cloud gas
although chemical effects may further complicate the interpretation
(\S~\ref{sec:obschem}). 
Thus, despite the enormous observational
effort, the dynamical importance of the magnetic field in clouds, especially
at the large scales, remains elusive. The current best guess is that
the magnetic field contributes in a non-negligible way to the energy balance
of molecular clouds. Whether it is a dominant player or a second
order effect cannot be decided
yet, and because of this, our understanding of the global
equilibrium of dark clouds is still incomplete.

\subsection{Equilibrium State and Star Formation}

The equilibrium state of dark clouds and the way they collapse 
under gravity to form stars is probably the most controversial
issue related to their nature. The uncertainties in the magnetic field
strength and the importance of the turbulent motions, together
with a possible revision of the lifetimes of the molecular
gas \citep[e.g.][]{har01}, have resulted in two opposed views of the global state of 
clouds.  Briefly, one view holds that clouds are close to equilibrium
and that their evolution toward star formation is approximately quasistatic.
The other view defends that clouds are dynamic objects that evolve 
and form stars in a crossing time. 
The limited space 
of this review cannot make justice to the number of issues involved
in this controversy or to the detailed position of each camp, which consists
of a number of authors working often independently. Here we will simply
review the main contentious points and mention a number of relevant
references as an introduction to the topic. We refer the reader to the
original papers and the reviews mentioned below for an in-depth view of 
the controversy.

The quasistatic view holds that  clouds are objects close to equilibrium,
due to their relatively long lives (at least 10 Myr according to 
the recent estimate by \citealt{mou06}, see also \citealt{bli80})
and their being gravitationally bound and close to virialized
\citep[e.g.,][]{lar81,mck99}. In this view, the equilibrium against
self gravity is provided by the magnetic field, which has a twofold
contribution toward stability. If the static component of the magnetic field
is strong enough to make the cloud subcritical, gravitational forces cannot
overcome magnetic forces as long as the field remains frozen into the matter
\citep{mes56,nak78}. Hydromagnetic waves,
in addition, can provide additional support and contribute to the supersonic
motions observed in molecular clouds \citep{aro75,gam96}. 
In this magnetically-dominated picture, molecular clouds can only evolve under 
gravity and form stars through the process of ``ambipolar diffusion,'' 
by which neutrals drift past the ions and the
magnetic field, which remain frozen to each other.
Through the action of ambipolar diffusion, cores of dense gas form
by gravitational contraction out of the initially subcritical medium.
When these cores have accumulated enough mass, they become supercritical 
and collapse to form stars \citep{mes56,shu87,mou99}.
Under typical cloud conditions, ambipolar diffusion is slow (several cloud 
dynamical times, see \S~\ref{sec:timescale}), and this slowness makes
cloud evolution prior to star formation occur quasistaticly. It also makes
star formation a rather inefficient process, in agreement with the 
low rate of stellar birth observed in the Galaxy \citep{zuc74}.

The more recent and opposite view of cloud evolution and star formation 
emphasizes the role of supersonic
turbulence and lack of equilibrium. Numerical simulations have shown
that magnetic turbulence decays in about a dynamical time
\citep{mac98,sto98,pad99},
so hydromagnetic 
waves cannot provide support against gravity without a source
of continuous replenishment. If in addition fields are weak 
enough so molecular clouds are supercritical \citep[e.g.,][]{nak98}, cloud 
evolution and star formation becomes a fast process that occurs in 
a crossing time \citep{elm00}. In this picture, clouds form from 
convergent flows, evolve, and dissipate
rapidly, without ever reaching a state of equilibrium, 
and with a typical molecular cloud lifetime of 3-5 Myr
\citep{bal99,har01,vaz03,har03}.   Molecular clouds are defined observationally by the detection of CO emission (as opposed to H$_2$ emission) and this lifetime refers to the age of the CO emitting cloud, which is associated with star formation.  Chemical models suggest that \Htwo\ formation is a pre-requisite to CO formation \citep{bergin_cform}, and lifetime estimates do not include any earlier phase when gas is predominantly molecular (e.g. \Htwo ) but CO has not formed in sufficient quantity for detectable emission.
Factors such as ram pressure, molecule and dust shielding, and gas-grain 
physics influence the timescales for both \Htwo\ and CO formation \citep{koy00,bergin_cform}.
But in any case, it is clear that substantial evolution ($\sim 10-20$ Myr) 
occurs in a pre-CO phase where the cold H~I  and \Htwo\ would be difficult 
to detect \citep{bergin_cform}.
  Formation of cloud structure in this scenario
is regulated by the stochastic action of turbulence, which produces
strong density perturbations through its fluctuating velocity field. Regions of
dense gas form at the
stagnation point of two convergent flows \citep{pad01}, although 
such a transient structure often disperses when the flows fade.
If the initial compression is strong enough, it may decouple 
from the medium and produce a dense core that subsequently collapses to
form stars. This process of creating structure through the interplay between 
gravity and turbulence is often referred as gravoturbulent fragmentation
\citep[e.g.,][]{kle04} and lies at the heart of
the dynamic picture of star formation 
(see the reviews by \citealt{mac04}, \citealt{lar03}, and \citealt{bal06}).

The controversy between the fast and slow modes of cloud evolution
and star formation is still ongoing and far from settled. Attempts
to determine cloud lifetimes using stellar ages, for example, have
not only not clarified the issue but have
led to their own controversy \citep[e.g.,][]{pal02,har03}. 
Chemical analysis of the gas can also provide an estimate of cloud timescales.  
Observations of cloud cores, for example, suggest compositions that are far from 
equilibrium and ages that are 
$\lesssim 3 \times 10^5$ yrs \citep[e.g.][]{pratap_tmc1, vdb_araa}, although
this chemical age might not reflect the true cloud age but rather refer to the 
last time 
the chemistry was reset by some dynamical event.  Studies of cold H~I in dark 
clouds, on the other hand, suggest longer timescales $> 3 \times 10^6$ yrs 
\citep{gol05}.
Measurements of the magnetic field, as mentioned before, remain 
ambiguous because the data populate the boundary between the critical
and subcritical regimes \citep{cru99}, and this is often interpreted
as proof of the magnetic field being dominant or
a minor player.  

The above summary of the current controversy illustrates how our understanding 
of dark cloud physics is still limited and far from complete.  The observations 
and theoretical work carried out over the last decade 
have clearly revitalized the 
field of cloud studies and challenged many ideas previously assumed correct. 
Unfortunately, they have not achieved yet a consensus view of how clouds form, 
evolve to form stars, and finally disperse.  It is therefore a challenge left 
for the next generation of cloud studies to explain the diversity
of observations and to develop a global view of clouds, from the large, 
tens-of-parsecs scale to the tenths-of-parsec size of the dense cores, 
to which we now will turn our attention.

\section{PHYSICS OF PRE-STELLAR MOLECULAR CORES}
\label{sec:physics}

Dense cores are localized density enhancements of the 
cloud material that
have been recognized for more than 20 years as
the likely sites of low-mass star formation in nearby dark
clouds like Taurus and Perseus.
Initial core studies, based on the
optical inspection of the Palomar plates together with molecular (especially 
NH$_3$) observations, showed that dense cores have typical sizes of 
0.1 pc and contain a few
solar masses of subsonic material at temperatures around 10 K and 
average densities
of a few $10^4$ cm$^{-3}$ \citep{mye83b,ben89}.
The correlation of dense core positions with the location of highly
embedded young stellar objects (YSOs) detected with the IRAS 
Far-Infrared (FIR) satellite
soon provided the proof that some dense cores are currently forming stars
or have done so very recently \citep{bei86}. The presence of an
embedded YSO has also led to the classification of
cores in two groups: starless and star-containing.
Strictly speaking, the distinction between the groups 
is only based on the threshold for detection of an embedded
YSO with current instrumentation, which was until
recently of the order of 0.1 $L_\odot$ for nearby clouds like
Taurus \citep{mye87}. The increase in the sensitivity of IR observations
brought by the Spitzer Space Telescope has led to
the discovery of a number of so-called very low luminosity objects 
(VeLLOs) in a few previously-thought
starless cores, forcing a revision of the core classification
\citep[e.g., L1014, see][]{you04}. Still, the distinction 
between starless and star-containing
cores seems a fundamental one, as it likely represents the 
separation between the pre-stellar and post-stellar phases of core 
contraction.
The study of starless cores, therefore, offers 
the best opportunity to determine the initial conditions of 
low-mass stellar birth.

Dark globules (often called Bok globules or Barnard objects)
are classically defined by their optical appearance as small,
roundish, and dark nebulosities \citep{bok47}.  They constitute
a less homogeneous class than the dense cores of dark
clouds, as they span sizes up to about 1 pc and masses of up 
to almost $10^3$ $M_\odot$, and have lower average densities 
\citep{bok77}. A number of small globules,
however, shares many characteristics with the dense cores, and are also
the formation sites of individual low-mass stars \citep{cle88,yun90}.
These dense globules may have a similar formation
mechanism as the dense cores, with the difference that globules 
do not lie embedded in a molecular cloud because they have been exposed by
an external event \citep[such as ionization from a nearby O star,][]{rei83}.
Like some dense cores, some globules seem starless and centrally concentrated,
and probably represent a phase prior to gravitational collapse. As we will see 
below, their isolated nature provides an additional advantage when studying 
the internal structure of pre-stellar objects \citep{alv01}.

The study of dense cores and globules prior to star formation
has progressed enormously during the
last decade. Earlier work lacked the angular resolution needed to study
their internal structure, and it therefore
concentrated on global properties like total masses and average
densities \citep[e.g.,][]{ben89,cle88}.
The progressive increase in resolution and sensitivity
of radio and IR observations has finally allowed a detailed view
of their interior.  Through a combination of 
techniques that trace the gas and dust components,
a new understanding of their internal structure 
has started to emerge. Crucial to this understanding
has been the realization that dense cores and globules, despite their
apparent simplicity, have a rather complex chemical composition
that is responsible for previous contradictions between 
observations made with different tracers. In this section we
review some of the recent results concerning the physical structure
of starless cores and globules, and the following section does the same
for their chemical properties. For the sake of economy, we 
will refer to both cores and globules simply as cores, and we will only make
an explicit distinction between the two when necessary. We will also use 
the term ``starless'' in a broad sense that includes what in the literature
is sometimes referred to as ``pre-stellar.'' 

\subsection{Density}
\label{sec:cores_dens}

\begin{figure}
\centerline{\psfig{figure=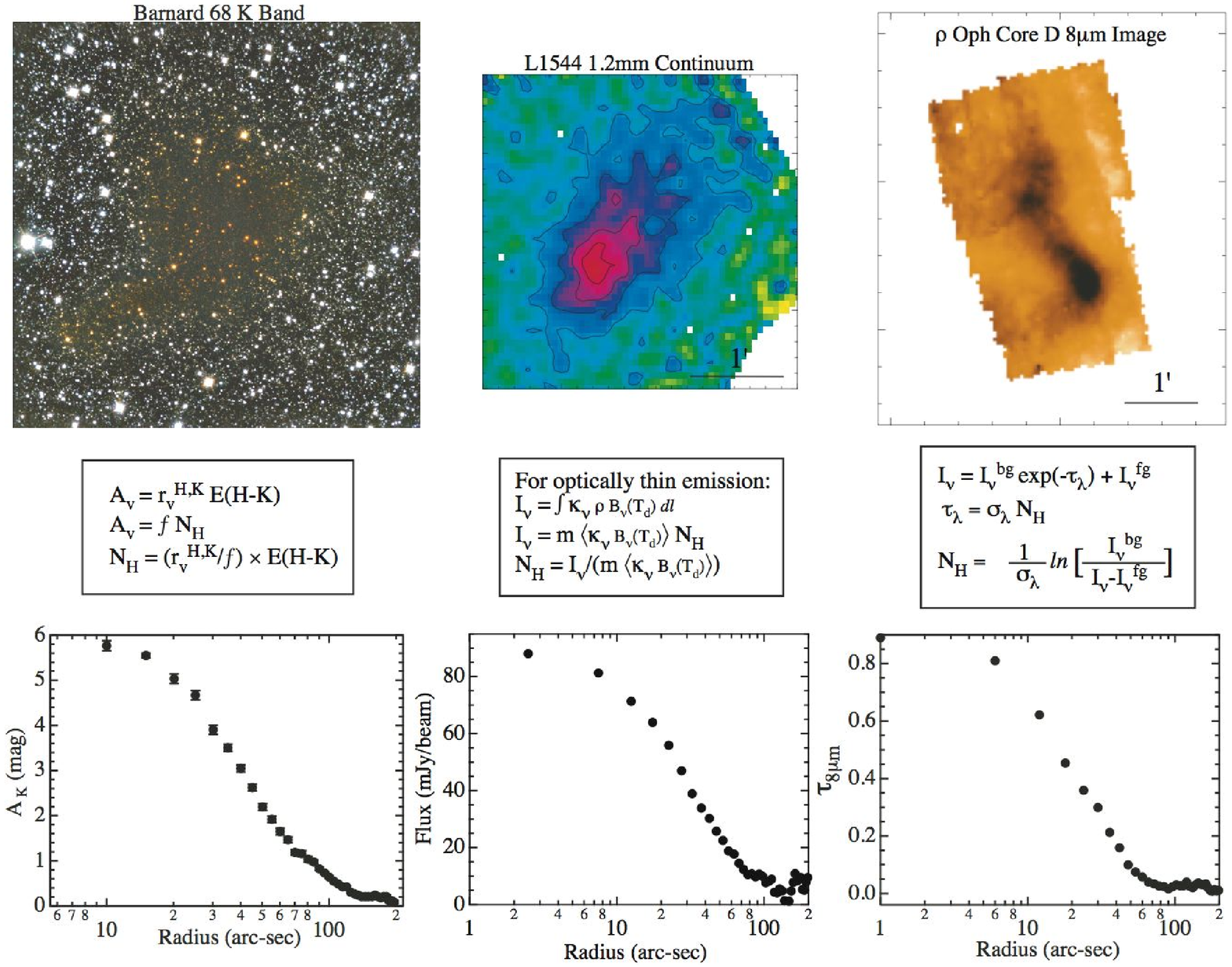,height=4in}}
\caption{Figure illustrating the various methods used to determine the 
column and volume density of starless molecular cores.  {\em Left Panels:} K-band 
image of Barnard 68 and plot of the derived A$_K$ with 10$''$ resolution as a 
function of core radius (taken from \citet{alv01} with A$_K$ vs. radius provided 
by C Rom\'{a}n-Z\'{u}\~{n}iga.  In this method the measured quantity is the H--K excess which is related to A$_V$ by the extinction law, parameterized by  $r_v^{H,K}$.  
A$_V$ is correlated to the hydrogen column from UV line measurements, parameterized by $f$
\citep{bsd78, rachford02}. 
{\em Middle Panels:}  1.2 mm dust continuum 
emission map and flux vs. radius for L1544 (taken from \citealt{war99} and 
\citealt{taf02}).   $\kappa_{\nu}$ is the dust opacity per unit gas mass, 
$\rho$ is the dust density, and $m$ the hydrogen mass (corrected for He).
{\em Right Panels:} 7 $\mu$m ISOCAM image and opacity vs. 
radius for $\rho$ Oph D (taken from \citealt{bac00}).   At 7 $\mu$m the emission from polycyclic aromatic hydrocarbons (PAHs) provides a bright background and the dense core appears in absorption.  In this method the absorbing opacity is related to the hydrogen column via the dust absorption cross-section, $\sigma_{\lambda}$.
\label{fig:denfig}}
\end{figure}

The density structure of pre-stellar cores is typically estimated through the
analysis of dust emission or absorption. Three main techniques have been used
so far: (1) near-infrared extinction mapping of background starlight, 
(2) mapping of mm/submm dust continuum emission, 
and (3) mapping of dust absorption against the bright 
galactic mid-infrared background emission. Each technique 
uses dust observations to derive a parameter that is proportional
to the gas column density, and then fits this parameter with a model 
density profile that assumes some simple geometry (usually spherical).
Fig.~\ref{fig:denfig} shows a summary of how each technique works, including
maps of representative cores (B68, L1544, and $\rho$ Oph D), 
the main equations used in the analysis, and the radial profiles of the 
dust-derived parameter to be fit with the density model.

Because of their different assumptions, the above techniques are sensitive 
in different ways to the properties of the dust and to the variation of these
properties inside the core. Uncertainties in the dust parameters propagate in
different manner and affect differently the final density determination
of each technique. 
The NIR extinction method, for example, is independent of
the dust temperature, but depends on the reddening law (using $r_v^{H,K}$ to
convert A$_K$ into A$_V$) and on the correlation between
A$_V$ and the gas column density, both of which may be variable  (see 
\citealt{rl85} and \citealt{bsd78} for typical ISM values).
The mm/submm dust continuum emission method, on the 
other hand, depends almost linearly on the dust temperature
and on the value of the dust emissivity ($\kappa_{\nu}$), two parameters that
are known to vary systematically as a function of core depth 
(see \ref{sec:cores_tdust} for more details). The mid-infrared 
absorption method is also subject to uncertainties. It is 
sensitive to fluctuations in
the emission from polycyclic aromatic hydrocarbons
(PAHs), which provides the bright background against which the
core is observed. It also depends on the somewhat uncertain 
relation between the
absorbing opacity and the hydrogen column density
(see \citealt{weingartner_draine} and \citealt{ragan_irdc}
for further details). Given the restricted use of each of these
methods (due to their difficulty and data 
requirements), no systematic inter-comparison between them 
has been carried out so far. As shown below, the methods 
agree in the qualitative characteristics of the density profiles,
but a quantitative comparison
of the inferred densities is still seriously needed. Without this,
we can only guess the relative uncertainties of our density
determination; current wisdom suggests that they 
are in the factor-of-2 range.

\begin{figure}
\centerline{\psfig{figure=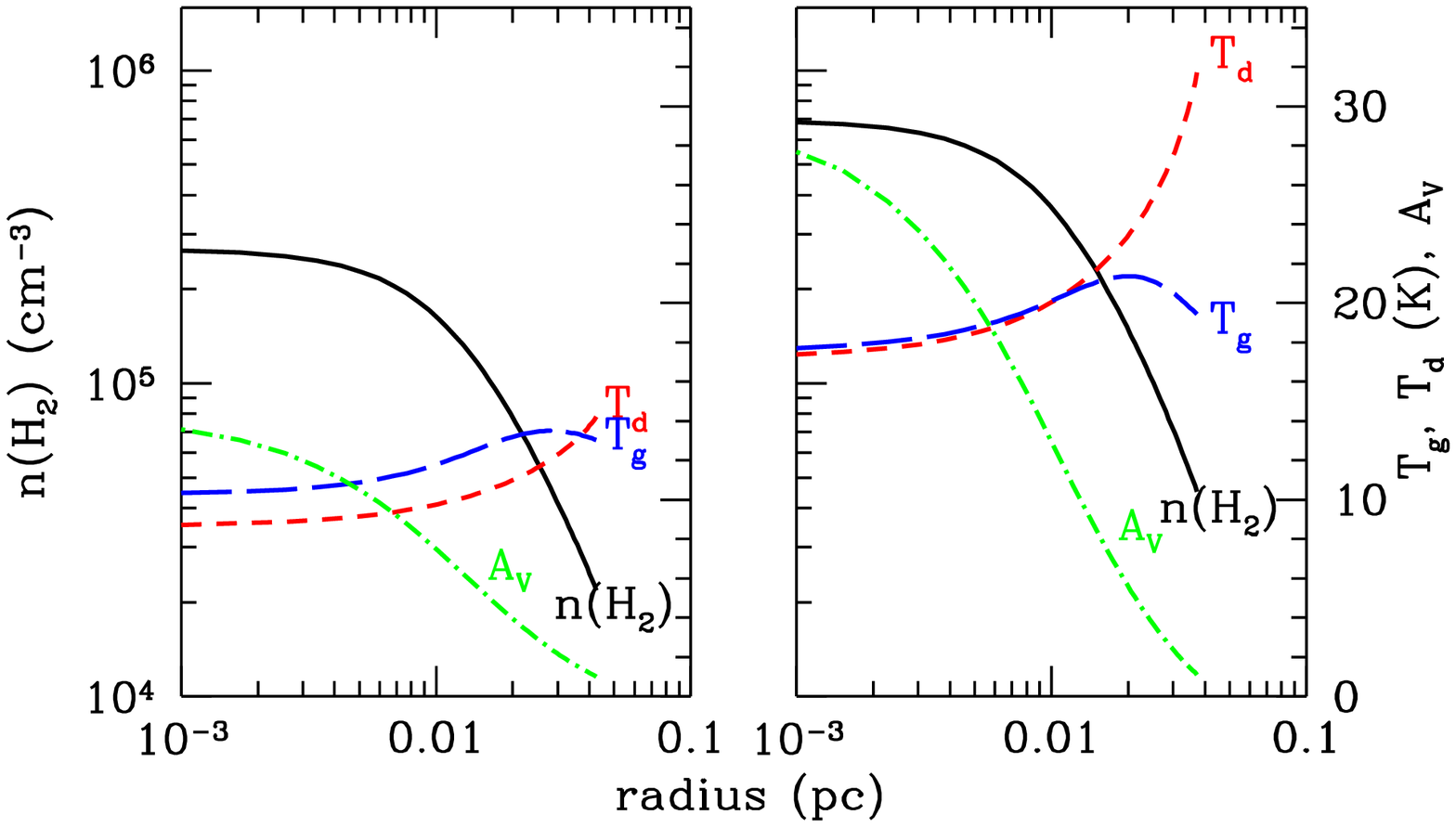,height=5in}}
\caption{{Left:}  Molecular hydrogen density, \nhtwo , gas  (T$_g$) and dust
(T$_d$) temperatures, and visual extinction (A$_V$) as a function of radius
for a marginally stable spherical cloud with M $= 1$ M$_{\odot}$ exposed to
the standard interstellar radiation field ($G_0 = 1$).
{\rm Right:} Same as {\em left} but for a marginally stable cloud
(M = 1.6 M$_{\odot}$) bounded by an external pressure ($p_{ext} = 10^6$ cm${-3}$ K)
and illuminated by an enhanced radiation field ($G_0 = 10$).
Figure taken from \citet{gal02}.
\label{fig:galli}}
\end{figure}

A main characteristic of the density profiles derived with the above
techniques is that they require a central flattening.
Pioneering work by \citet{war94} showed that the radial profiles of 
submillimeter continuum emission for a number of cores could not be 
fitted using single power-law density distributions (assuming a constant dust
temperature). At least two power laws were needed to fit the
data, and the inner power law was always flatter than the outer 
one, which was usually close to $r^{-2}$. 
Follow up work with higher sensitivity 
has confirmed these early results for an increasing number of 
objects, and it has shown that as a rule, the density gradient
of a core is flatter than $r^{-1}$ within radii smaller than
2500-5000 AU, and that the typical central density of a core is  
$10^5$-$10^6$~cm$^{-3}$ \citep{and96,war99}.
Although a dust temperature drop toward the core center 
is expected from thermal equilibrium considerations \citep{eva01}, 
the effect seems not strong enough to explain
alone the flattening found in the continuum data. Observations using 
extinction techniques, which are insensitive to temperature, also
find that a central flattening of the density profile is needed
to fit the data. \citet{bac00}, for example, 
have modeled MIR ($7 \mu$m) absorption profiles from a sample of
nearby starless cores observed with ISOCAM, and they have found that
a decrease in the slope of the density distribution is
required to fit the emission in the central 4000-8000 AU.
\citet{alv01} have mapped in detail the NIR extinction
from the B68 globule and achieved an almost perfect fit with an
equilibrium isothermal (Bonnor-Ebert) density profile that has a 
central flattening within $\sim$5000 AU.  Density profiles comparable to those inferred from observations are shown in Fig.~\ref{fig:galli}.

A number of density laws with central flattening have been used to fit 
the observations of dense cores. The dual power law used initially
by \citet{war94} is illustrative of the need for inner
flattening, but presents an unphysical discontinuity
in the derivative that results in 
a pressure jump. A simple alternative is the 
softened power laws used by \citet{taf02}, and a more physical
model is the truncated isothermal (Bonnor-Ebert, BE) sphere
\citep{ebe55,bon56}, that often (but not always) provides
a good fit to the data. Examples of good BE fits 
include those to the extinction profiles of 
a number of Bok globules \citep{alv01,kan05}
and dense cores \citet{bac00}.
Other physical models, on the other hand, seem to be ruled
out by these observations. The logotropic
density profile of \citet{mcl96} is too
flat at large radii ($r^{-1}$) to fit most cores \citep{bac00},
while the Plummer-like profile \citep{whi01} seems asymptotically 
too steep ($r^{-4}$). Less clear is the situation of magnetic 
field-dominated models. These models naturally produce
equilibrium configurations with flattened density profiles 
and approximately $r^{-2}$ power-law behavior at large radius
\citep[e.g.,][]{mou76,tom88,liz89}.
\citet{bac00} find that some of these configurations 
provide adequate fits to their data,
but that they require magnetic fields of the order of
50-150~$\mu$G, significantly larger than commonly observed
(\S~\ref{sec:cores_bfield}). They also predict oblate geometries, 
which seems to contradict some observations \citep{mye91b}. 

The high quality of some BE fits raises the question of whether
cores are truly isothermal spheres solely supported by thermal
pressure and confined by an outer medium. 
Several lines of evidence suggest that although thermal
pressure is an important ingredient in core structure,
it is not the only one, and that cores are more complex
entities than simple BE spheres.
In the first place, cores are seldom spherical, but they appear
in the sky as elliptical objects with axial ratios typically of 2:1
\citep{mye91b}. Whether cores are prolate, oblate, or triaxial
is still a matter of some debate \citep{mye91b,ryd96,jon01}, but the
lack of spherical symmetry already suggests either the presence of a
non symmetrical force component or the lack of perfect equilibrium. 
In addition, the observed density contrast between the core center 
and its edge often exceeds the maximum value of $\approx 14$ allowed
for stable BE spheres \citep{bac00,eva01}, and it seems unlikely that
such an unstable configuration can be realized in nature 
if thermal pressure is the only balancing force. 
Finally, the pressure required by the BE fits usually
exceed the thermal (plus subsonic non-thermal) pressure available from 
the gas by a factor of a few, again indicating the presence of an additional force
component or the lack of equilibrium \citep{taf04,kir05}. 

The 
BE-like profiles found when fitting observations must therefore 
result from a more complex physics than just thermal hydrostatics.  Indeed, \citet{mye05} and \citet{kan05} have shown that these profiles 
can be reproduced with a variety of geometries and even under early conditions 
of gravitational collapse. 
 Even models of supersonic turbulence can
produce density condensations that have in projection
the appearance of BE profiles \citep{bal03}, but their expected velocity
structure is inconsistent with the prevalence of subsonic
cores in dark clouds like Taurus \citep[e.g.,][Tafalla et al. 2007,
in preparation]{jij99}. The ultimate origin of the observed 
core density profiles remains therefore unexplained.

\subsection{Temperature}

The kinetic temperature of the dust and the gas components in a core
is regulated by the equilibrium between heating and cooling. At typical
core densities ($> 10^4$~cm$^{-3}$), the gas and dust start
to couple thermally via collisions \citep[e.g.][]{gol78,bur83}, and the
two temperatures are expected to converge; whether they equalize or not depends
in part on the exact density of the core and on the strength of the external
radiation field \citep{gal02}. In this and the next sections we discuss
recent progress in the determination of the dust and gas temperatures
of cores. As the two temperatures are determined using different
observational techniques and are therefore subject to different 
uncertainties, we separate their discussion. We start with the dust.

\subsubsection{Dust Temperature}
\label{sec:cores_tdust}
The dust temperature is physically determined
by the balance between the heating by the interstellar radiation field
(ISRF) and the cooling by thermal radiation from the grains in the FIR \citep{mat83}.
To estimate this temperature observationally, color measurements of the 
dust in the submm or FIR are commonly used, sometimes complemented with 
NIR extinction measurements.
A major uncertainty in this work is our limited understanding
of the optical properties of the dust and, in particular, the value of the
dust emissivity and how it changes as a function of density. Theoretical
models of dust evolution predict that as the density increases, the dust grains
become coated with ice and coagulate to form fluffy aggregates,
changing their emissivity in the process \citep{oss94}.
Determining dust temperatures
from observations therefore implies resolving simultaneously for their
optical properties, in particular the emissivity ($\kappa$) and its 
wavelength dependence (commonly parametrized as $\nu^\beta$).

Large-scale studies of dust temperature show that the grains
in starless cores are colder than in the surrounding lower-density medium,
as expected from the attenuation of the ISRF deep inside the cloud.
From ISO FIR (170 and 200~$\mu$m) observations toward 
the vicinity of a number
of dense cores, \citet{war02} found a prevalence of flat or decreasing
temperature gradients with cloud temperatures of 15-20~K and core 
values of 10-12~K. Because of the 1 arcminute resolution 
of these observations (0.04 pc at the distance of Taurus), 
the core value represents only an average temperature
over the densest gas. Also using low-resolution ISO data and a similar 
analysis, \citet{toh04} and \cite{del05} have derived inward temperature gradients
that reach central values of $\approx 12$~K for a number of Taurus cores, 
while for the denser L183 core a value close to 8~K has been measured 
\citep{leh03,pag03}. These temperature estimates, however, 
ignore the effects of grain evolution and
assume a constant dust emissivity, while observations are 
beginning to show that the emissivity does in fact vary.
FIR observations of a filament in Taurus by the PRONAOS 
balloon-borne experiment suggest that at the highest densities, the population
of smallest grains disappears and the submillimeter emissivity increases by
a factor of about 3, as expected by models of dust coagulation into fluffy 
grains \citep{ste03}. These observations also suggest inner dust temperatures 
around 12~K, although the 3-arcminute resolution sets an important limitation.

The trend of decreasing dust temperature with depth 
observed in the large scale maps 
is expected to continue inside the dense cores, but the exact
shape of the resulting gradient is still a matter of some uncertainty. 
Initial models by \citet{eva01} and \citet{zuc01} suggested 
sharp core edge-to-center
drops in temperature of about a factor of 2 (an example is provided in 
Fig.~\ref{fig:galli}).  However, these models either assumed that
cores are directly exposed to the ISRF (correct for Bok globules
only), or use a simple attenuation law for cores that are embedded
in clouds. \citet{and03} and \citet{sta03} have argued that the effect of
the surrounding cloud is to flatten the temperature gradient, so that 
at the center of an embedded core the gradient is less pronounced 
\citep[see][for 2D and 3D radiative transfer models]{sta04,gon04}. 
Irrespective of the models, however, central dust temperatures close 
to 7-8~K are predicted for typical dense cores, with the
exact value depending on the dust central density and the presence
or not of a surrounding cloud. 

\begin{figure}
 \centerline{\psfig{figure=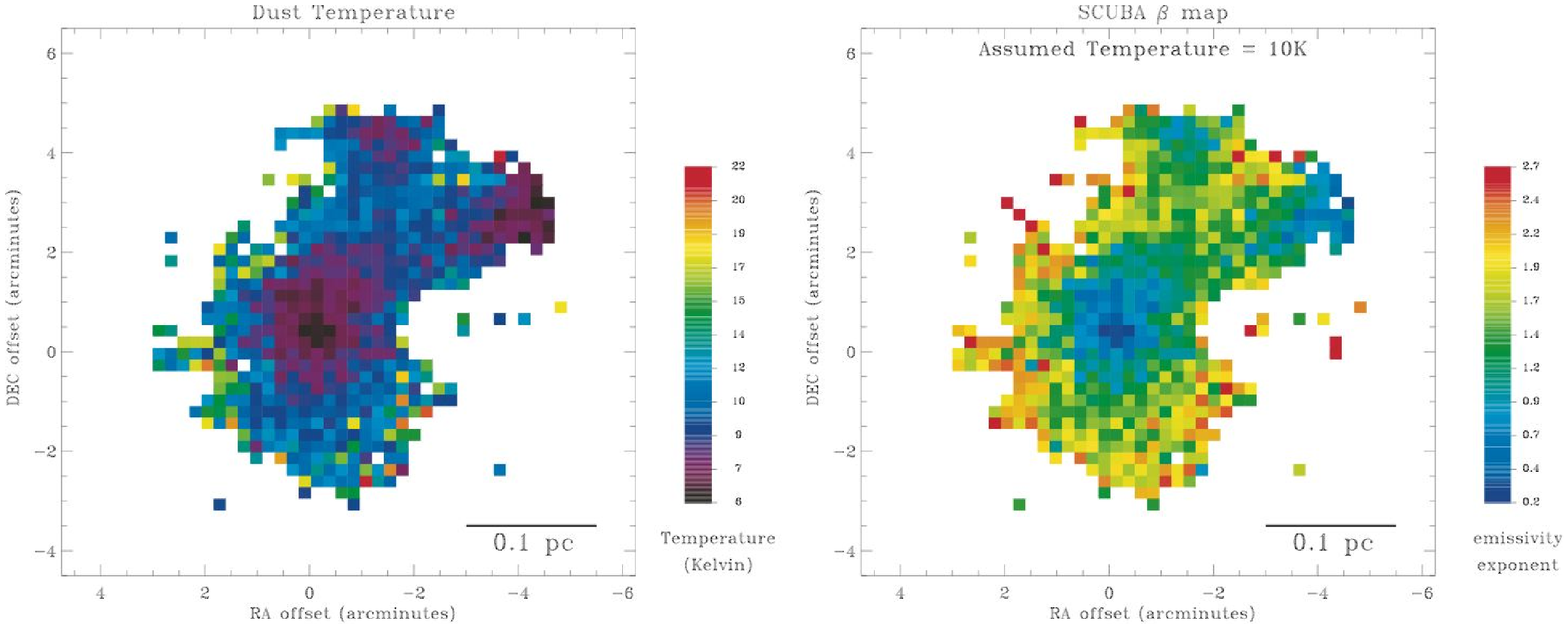,height=2.3in}}
 \caption{{\em Right:} Map of dust temperature in TMC-1C derived from 450 and 850 $\mu$m SCUBA maps assuming dust emissivity index $\beta$ = 1.5. {\em Left:} Map of dust emissivity index, $\beta$ in TMC-1C assuming a constant temperature of 10 K.   Taken from 
 \citet{sch05}. \label{fig:schnee}}
\end{figure}

From an observational point of view, 
possible gradients in the dust emissivity still limit our
estimate of the temperature profiles.
\citet{kra03} have combined submm
emission and NIR extinction data to study several starless cores in IC~5146.
With an angular resolution of 30 arcseconds, these authors find that
the dust temperature is typically 12~K and close to constant inside the cores,
while the surrounding, lower extinction material has a higher 
value of about 20~K. They also find that the dust emissivity increases 
systematically by a factor of 4 between 20 and 12~K, consistent
with models of dust evolution \citep[e.g.,][]{oss94}.
Using similar technique and
angular resolution, \citet{bia03} have also found an emissivity gradient in 
the B68 globule, together with an outer dust temperature of $14\pm2$~K and
an inner temperature drop of a factor of 1.5.
\citet{sch05}, on the other hand, have studied 
the submm emission from TMC-1C with $14''$ resolution and found a central gradient
in the submm color. As illustrated in Fig.~\ref{fig:schnee}, this color
gradient can result from either a central temperature drop
to about 7~K or a systematic change in the dust emissivity index (or both).
Clearly more data and modeling are needed to characterize the existing gradients
in dust conditions inside dense cores.
Future multiwavelength mapping in the FIR with the Herschel
satellite offers the best hope to finally disentangle the changes in 
temperature and optical properties.

\subsubsection{Gas Temperature}
\label{sec:cores_tgas}
Like the dust temperature, the temperature of the gas is determined
by the balance between heating and cooling. For the 
lower density cloud gas, heating mostly occurs through
ionization by cosmic rays, while the cooling 
is mainly due to line radiation from molecules, especially CO \citep{gol78}.
At the high densities of cores (more than a few $10^4$ cm$^{-3}$),
gas-dust coupling via collisions becomes important, and this
process will additionally heat or cool the gas depending
on the difference between the gas and dust temperatures \citep{bur83}.
Also at these densities, and at typical core temperatures, 
CO and other molecules start to disappear 
from the gas phase due to their freeze out onto the dust grains
(\S~\ref{sec:obschem}), diminishing the ability of the gas 
to cool down via line radiation. The detailed study of thermal balance in dense
cores by \citet{gol01} shows that the gas-dust coupling, together
with the high efficiency of the dust radiative cooling, easily compensates
the disappearance of molecular coolants, and keeps the 
temperature of the gas deep inside the core at rather low values, comparable to
those of the dust component. 
At the lower-density core edges, the gas and dust are thermally decoupled 
and the gas becomes warmer than the dust due to photoelectric heating.
The temperature difference between the two components depends on the
strength of the external UV field (see Fig.~\ref{fig:galli}).

Keeping the gas cold and at
a close-to-uniform temperature
has important implications for the mechanical
equilibrium of cores, as the gas component is the main
contributor to the thermal pressure. \citet{gal02}
have studied the equilibrium structure of a pressure-supported
dense core with a realistic treatment of the gas and
dust cooling. Although the resulting gas temperature distribution 
is not exactly constant, its gradient is shallow enough 
to require for stability a density profile which is very close
to the Bonnor-Ebert solution (see Fig.~\ref{fig:galli}). Thus,
if cores were supported by thermal pressure only, the BE
model would be consistent with the internal thermodynamics.

To measure the temperature of the gas component in clouds and cores,
the level excitation of simple molecules like CO and NH$_3$ is
commonly used
\citep{evans_araa}. CO and its isotopologues
have small dipole moments, so their lower energy levels thermalize at 
relatively low densities of the order of few $10^3$ cm$^{-3}$
\citep{spi78}, typical of the extended gas in dark clouds.
Temperatures derived using CO in regions with no star formation
indicate gas temperatures of 10-15~K, with a possible
increase towards the lower density gas near the cloud edges \citep{sne81}.
Temperatures for the denser gas in cores are better determined using
the metastable inversion levels of ammonia (NH$_3$), which can be observed
simultaneously at a wavelength of about 1.3 cm \citep{wal83}. NH$_3$,
in addition, appears to rise in abundance in the densest ($n_{\rm H_2} > 10^4$ \cc ) regions
(\S~\ref{sec:obschem}), so it tends to selectively trace dense core gas. Low angular resolution NH$_3$ observations
of dense cores typically indicate gas temperatures of about 10~K with a rather
narrow range of variation for clouds like Taurus, Perseus, and Ophiuchus
\citep{myers83,ben89,jij99}. These low-resolution 
measurements represent core-wide averages of the temperature,
and are in reasonable agreement with the theoretical expectations mentioned 
before. Due to the relatively low angular resolution of the NH$_3$ 
observations (caused by the low frequency of the transitions), few studies
of the radial dependence of the gas temperature inside 
dense cores have been made. \citet{taf04} have studied the temperature
profiles of two Taurus-Auriga starless cores, L1498 and L1517B, using 
NH$_3$ observations with a resolution of $40''$ (0.027~pc). In both cores, 
the temperature remains constant over the central 0.1~pc with a value of 
about 10~K. Higher resolution NH$_3$ observations 
of the denser starless core L1544 seem to indicate
a decrease of the gas temperature toward
the innermost parts of the core (Crapsi et al. in preparation).
For the starless Bok globule B68, on the other hand, \citet{ber06} have found 
an opposite effect: while the inner core traced by NH$_3$ is at about 10~K,
the surrounding layers, traced with CO isotopologues, are slightly cooler
(7-8~K). This central warming may result from a weakening of the gas-dust 
coupling at the highest densities due to grain coagulation. Given this
diversity of results, 
further observations of the temperature gradients inside 
dense cores are needed to clarify the origin of
these significant, but relatively small (20\%), temperature differences.

\subsection{Velocity Structure}
\label{sec:cores_veloc}

\begin{figure}
\centerline{\psfig{figure=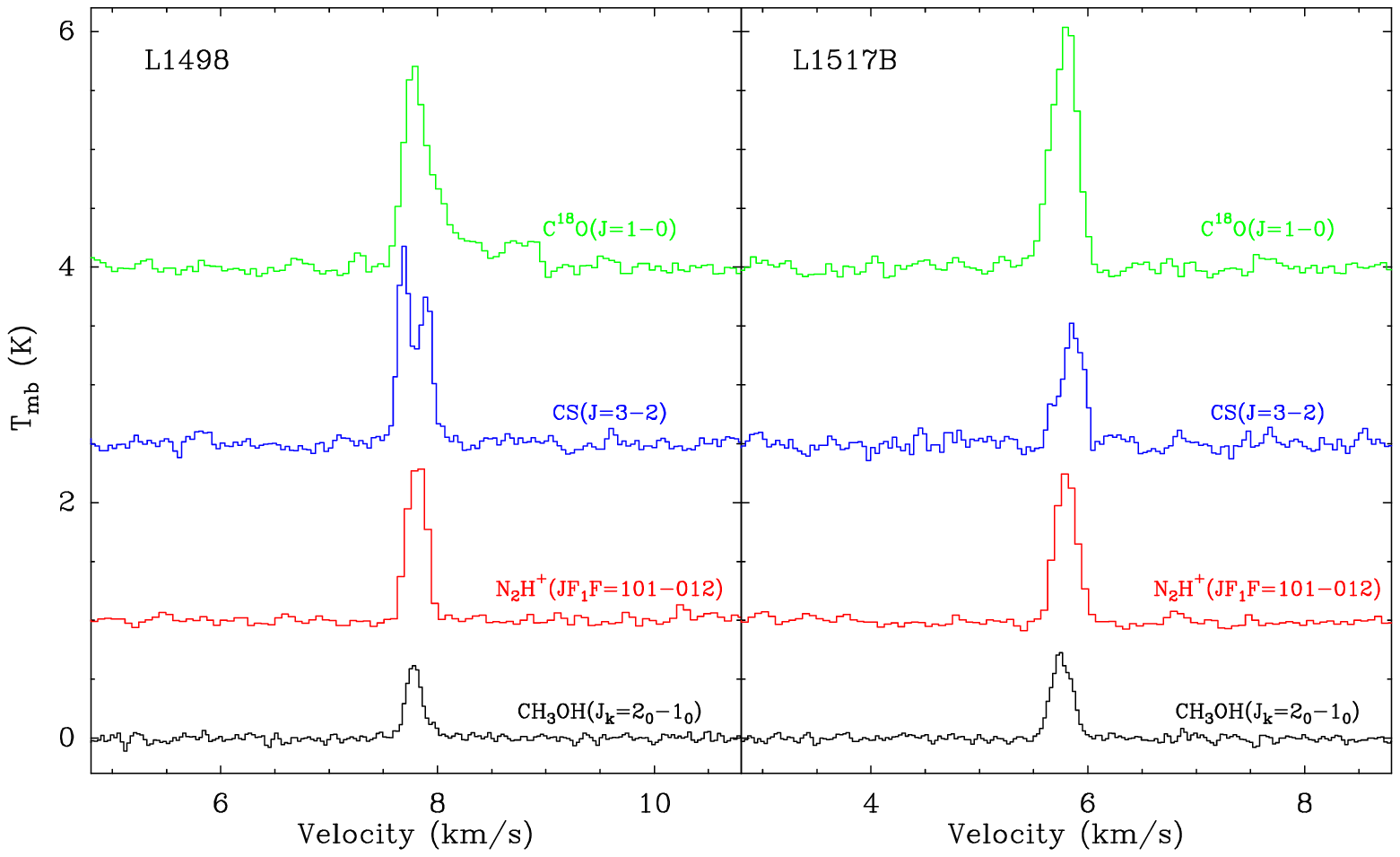,height=3.5in}}
\caption{Sample spectra from the starless cores L1498 and L1517B
in the Taurus-Auriga dark cloud illustrating the variety of 
shapes and intensities commonly observed. The C$^{18}$O spectra
trace both dense and low-density gas (due to the low dipole moment
of the molecule), and often present additional velocity components
from unrelated gas that moves at velocities of the order
of 1 km s$^{-1}$ (red gas in L1498). The optically thick CS spectra 
are sensitive to the relative motions of the core and its
envelope thanks to the presence of self-absorptions, that commonly
indicate inward (L1498) or outward (L1517B) motions at the level
of 0.1 km s$^{-1}$. The optically thin N$_2$H$^+$ and CH$_3$OH
lines trace different regions of the core depending on their 
chemistry, and their narrow linewidths reveal the subsonic
nature of the gas internal motions. Data from \citet{taf04} and
\citet{tafalla06}.
\label{fig:l1498_l1517b}}
\end{figure}

In contrast with the supersonic velocity fields characteristic of 
the clouds as a whole, dense cores have low velocity internal motions.
Starless cores in clouds like Taurus, Perseus, and Ophiuchus systematically
present spectra of core-tracing species (e.g., NH$_3$) that have
close-to-thermal linewidths, even when observed at low angular resolution
\citep{myers83,jij99}. This indicates that the gas motions inside the
cores are subsonic, either in the form of random turbulence or as part of
more systematic global patterns (Fig.~\ref{fig:l1498_l1517b}).

Despite the slow speeds involved, the internal motions 
of cores are in general complex, with no single element identified as
dominant. The excess in linewidth over the thermal value
has commonly been interpreted as resulting from turbulent motions, although
as we will see below, other velocity patterns like infall 
could contribute to the line broadening. Even in this 
interpretation, turbulent motions, by being subsonic, contribute 
less to the gas pressure than the thermal component, so they
represent a minor contribution to the core support
\citep{myers83,goo98,cas02,lad03,taf04}.
Studying the spatial distribution of turbulent motions in
cores, \citet{goo98} have found that they deviate from the
linewidth-size relation seen at large scales 
(\S~\ref{sec:cloud_veloc}, \citealt{lar81}).
Instead, the non-thermal component of the 
NH$_3$ linewidth in a core is approximately constant inside
a ``coherence'' radius of about 0.1~pc
\citep[see also][]{cas02,taf04}. 
This coherence radius is close to the 0.04~pc radius found
to separate the binary and clustering regimes in Taurus and interpreted 
as the length-scale at which the cloud 
self-similar behavior breaks down 
(\S~\ref{sec:cloud_veloc}, \citealt{lar95}).
It is therefore tempting to
suggest that the coherence radius is the scale
at which turbulence loses its grip on cloud motions and allows the
formation of thermally dominated cores.

A rotation components is 
expected in any object that forms by the gravitational contraction of 
galactic material \citep{spi78}. From the study of the spatial
changes in the velocity centroid of lines, however,
it has been known for some time that dark clouds as a whole are not rotating
rapidly \citep{hei76,arq86}. Observations of dense cores and globules 
reveal a similar lack of fast rotation in the dense gas.
\citet{goo93} analyzed the rotation patterns of
43 dense cores as derived from observations of the NH$_3$(1,1) line,
and estimated that the typical ratio between rotational
and gravitational energies is only 0.02. These authors also found no correlation
between the rotational axis and the orientation of core
elongation, which is again indicative of rotation being only a
minor element in core support. 
From higher resolution N$_2$H$^+$ observations of an equivalent sample
of cores, \citet{cas02} have derived similarly low values of the rotational
energy, and found that the point-to-point velocity gradients inside
the cores are much more complex than expected for simple solid-body
rotation. Rotation is therefore only one of the ingredients
of core kinematics.

The slow rotation rate of dense cores is in fact remarkable.
\citet{arq86} and \citet{goo93} have shown that 
the specific angular momentum $j$ in a core or a cloud scales with
its radius $R$ approximately as $R^{1.6}$. Clouds of 1 pc radius have
typical $j$ values around $10^{23}$ cm$^2$ s$^{-1}$, while 
typical cores 
(0.05 pc radius) have $j$ values of about  $10^{21}$ cm$^2$ s$^{-1}$.
If this trend of decreasing $j$ reflects
the evolution of specific angular momentum as the cloud gas 
undergoes contraction, an efficient mechanism of angular momentum
removal is required to reproduce the observations, and in
particular, to explain the low $j$ values of cores. Magnetic coupling
between the contracting gas and the lower density environment
has been classically favored as the mechanism for transferring 
angular momentum from a contracting core to its surroundings 
\citep{mes56,mou79}, although 
an alternative interpretation in terms of gravoturbulent fragmentation
has recently been proposed \citep{jap04}. The low $j$ value of
cores is in fact comparable to or slightly larger than
the value inferred for protostellar envelopes, T Tauri disks, and wide binaries,
suggesting that the later phase of gravitational contraction
of the core
occurs with almost constant angular momentum \citep{bod95}. 
Of course, a
final phase of angular momentum redistribution via disk
transport or tidal interactions between stars is still needed
to explain the low specific angular momentum found in individual
T Tauri stars \citep[e.g.,][]{shu87,bod95,lar02}.

Inward motions can also contribute to the dense core
kinematics. They are inferred from the observation of
optically thick, self-absorbed lines of species like CS, H$_2$CO, or 
HCO$^+$, in which low-excitation foreground gas absorbs part
of the background emission. In a core with inward motions,
the foreground gas moves toward the background, so 
the absorbing material is red-shifted with respect to the emitting
layers. The resulting spectrum is 
a self-absorbed profile with a red-shifted dip and a 
relatively brighter blue peak (see CS J=3--2 in Fig.~\ref{fig:l1498_l1517b}),
a pattern often called an
infall asymmetry \citep{leu77,mye96}. In a starless core, this spectral feature
has been first identified towards  L1544, for which an average inward velocity of about
0.1 km s$^{-1}$ was inferred \citep{mye96,taf98,wil99,cas02b}. A follow up 
single-pointing survey of 220 starless cores by \citet{lee99} has found a
statistically significant over-abundance of inward motions
with typical velocities of 0.05-0.1 km s$^{-1}$, indicating
that subsonic contraction is a common feature of core kinematics
\citep[also][]{gre00}. The spatial distribution of
these motions is typically large, 0.05-0.15~pc, and comparable
to the observed size of the cores \citep{taf98,lee01}. 
 Such a large large extent could result either from the 
gravitational collapse of a core having a centrally condensed initial 
state \citep{mye05} or be associated
with the process of core condensation \citep[see][for a review]{mye00}.

Additional velocity patterns have been proposed to explain the
observations of molecular lines in different cores. \citet{lad03}
have found an asymmetric velocity field with both inward and
outward motions in the B68 
globule, and they have interpreted it as resulting from small 
amplitude pulsations of the core outer layers. Such a pattern seems
in fact consistent with the close to Bonnor-Ebert density structure
found in this object \citep{ket06}. \citet{taf04} have proposed
that residual internal motions in L1498 and L1517B may originate from
asymmetric contraction, as the motions are correlated with asymmetries
in the pattern of depletion of species like CO and CS. Finally, by comparing 
the velocity centroids of N$_2$H$^+$ and CO isotopologues in a sample
of 42 cores, \citet{wal04} have concluded that the relative velocity between
the dense cores and the ambient medium is low ($\le 0.1$~km s$^{-1}$).
This implies that newly born stars are not expected to move 
appreciably from their natal cores.

\subsection{Magnetic Field}
\label{sec:cores_bfield}

Estimates of the magnetic field
in starless cores are mainly obtained from Zeeman splitting measurements
and from polarization observations of the mm/submm
dust continuum. Zeeman measurements have the disadvantage of a low
angular resolution and the use of tracers sensitive to molecular 
depletion, so they are prone to miss the densest part of the cores.
On the other  hand, they are unique because they
provide a direct estimate of the strength of the (line of sight 
component of the) magnetic field, $B_{los}$. From  OH 
Zeeman measurements of the L1544 core with the Arecibo telescope,
\citet{cru00} have estimated a line of sight strength of about 11~$\mu$G
averaged over a beam of 3 arcmin (about 0.1~pc), in good agreement with 
the prediction from a previous model of ambipolar diffusion \citep{cio00}.
These observations, however, are more sensitive to the large-scale
cloud gas than to the core material because of the the low dipole moment and the 
chemical distribution of the OH molecule in clouds (\S~\ref{sec:obschem}).
Zeeman measurements using tracers of dense gas have mostly produced
negative results, like $B_{los} \le 100$~$\mu$G in L1498 
\citep[CCS data,][]{lev01}
and $B_{los} \le 15 \mu$G in TMC-1 \citep[C$_4$H data,][]{tur06}. 
\citet{shi99}, on the other hand, have reported an estimate of
$160 \pm 42 \mu$G in the young starless core L1521E from CCS observations.

Submm polarization measurements provide a direct estimate of the 
orientation
of the (plane of the sky component of the) magnetic field.
They can also be used to estimate indirectly the strength of the field
by comparing the spatial dispersion of its orientation and the
turbulence level of the gas \citep{cha53}. As mentioned in 
\S~\ref{sec:cloud_bfield}, these measurements often
lose sensitivity towards the brightest, highest density regions,
where the polarization fraction drops with intensity
\citep{mat00,hen01,wol03}, which again makes the estimates more
sensitive to the field in the outer parts of the cores. 
Because of the combination of weak submm emission of the starless cores 
and a low polarization fraction ($\sim 10$\%), 
these observations are challenging and have only been carried out
in a reduced number of objects. For cores L1544, L183, and L43,
\citet{war00} and \citet{cru04} have found polarization
vectors that are fairly uniform but not aligned with the 
minor axis of the cores by about $30^\circ$, as it would have been expected by
ambipolar diffusion models (but see \citealt{bas00} for 
an alternative view if the cores are triaxial). From the dispersion
of the polarization vectors and the turbulence level of the gas, these 
authors estimate plane-of-the-sky intensities between 80 and 160~$\mu$G.
Strictly speaking, these field intensities make the cores supercritical
(i.e., with a magnetic field too weak to balance gravity). However, 
the inclusion of a geometrical correction factor moves the cores to the
critical or slightly subcritical regime, suggesting that
the field may be dynamically important \citep{cru04}. Lower field values 
of 10 and 30~$\mu$G have been estimated towards the cores L1498 and L1517B,
respectively, indicating that these cores may be supercritical
by factors of about 2 even after geometrical correction \citep{kir06}.

As with the large scale magnetic field measurements
(\S~\ref{sec:cloud_bfield}), the observations of cores provide an 
ambiguous answer to the question of whether magnetic fields
dominate or not the equilibrium of dense cores. 
Clearly more observations of both the Zeeman effect and the 
dust polarization are needed to clarify the 
role of magnetic fields in the process of core and star formation.
The close to critical values observed so far, however, suggest
that unless observations have been systematically biased in one
or the other direction, cores appear to be close to the point of
criticality to within a factor of about 2. Explaining
this apparent coincidence should therefore be a main requirement
to any theory of core formation.

\section{CHEMISTRY OF PRE-STELLAR MOLECULAR CORES}

\subsection{Background}
The presence of gas phase chemical inhomogeneities within molecular clouds has been known for some time.   In particular, spatial differences in emission morphologies between carbon chains (e.g. \HCthreeN , \CtwoS ) and nitrogen-hydrogen molecules (\NHthree , \NtwoHp ) on scales of 0.05--0.2 pc were described \citep{suzuki92, pratap_tmc1}.   Surveys of numerous cores in Taurus revealed larger emission sizes and velocity linewidths for CS when compared to \NHthree , despite similar line centroids \citep{zhou89}.
These differences have generally been attributed to the slower timescales to activate the nitrogen chemistry as opposed to the carbon chemistry, or perhaps due to some dynamical cycling of material that can enhance the carbon chemistry \citep[see][for a discussion of general chemistry]{vdb_araa, langer_ppiv}.

In this regard two low-mass cores, TMC-1 and L134N, were isolated and studied as template objects because of their rich chemistry and due to the fact that they are unassociated with star formation activity
 \citep{swade89, hirahara_tmc1, pratap_tmc1}.
Over one thousand papers have been written for these two objects alone.
 In general chemical studies computed average line of sight abundances using a technique outlined in \citet{igh}.   H$_2$ does not emit appreciably at the cold temperatures associated with these objects (T $\sim$ 10 K).  Therefore optically thin isotopologues of CO, generally \CeiO , were used as surrogates to trace the H$_2$ column density with previously calibrated abundances 
\citep[e.g.  $x$(\CeiO ) $\sim 1.7 \times 10^{-7} \equiv$ n(\CeiO )/n(\Htwo );][]{flw}.  These studies strongly aided the interpretation that chemical reactions between ions and molecules dominate the gas-phase chemistry
\citep{herbst_chemrev}.

In parallel to gas-phase chemical studies, the fundamental vibrational modes of solid state molecules such as \HtwoO\ and CO were detected in the ISM \citep{gillett_h2o, lacy_co}.  Many studies of molecular ices frozen on the surfaces of dust grains have been conducted towards embedded sources which provide a strong infrared background to study foreground ices \citep[e.g.][]{vd_araa}. To provide important insight on ices in cold gas, observations of bright field stars located behind molecular cloud material (with extinctions $\leq$ 20 mag) are used as candles that probe
material remote from embedded sources \citep{whittet1988}.   Icy mantles coating grains in ambient gas are dominated by \HtwoO\ ice with substantial amounts of CO and \COtwo\ at a level of $\sim$25\% of the water ice abundance in each case \citep{gibb_iso}.    

There exists a threshold extinction below which the ice features are not seen, implying that grains are not mantled at low extinction.  This threshold varies for each ice component and is different between various clouds.   In Taurus, the water ice threshold is A$_V^{th}({\rm H_2O}) = 3.2^m\pm0.1^m$; a comparable threshold is observed for CO$_2$ ice but for CO A$_V^{th}({\rm CO}) = 6.8^m\pm1.6^m$  \citep{whittet01, ber05, whi07}.  The threshold value presumably is related to the formation mechanisms for ices and in Taurus the relatively low water ice threshold may imply that the water ice forms at a low density were atomic H is available.
 These observations further demonstrate that ices constitute a substantial fraction of the available carbon and oxygen \citep{vdb_araa}. Thus CO (and its isotopologues) is not a reliable tracer of  the H$_2$ column when using previously calibrated abundances.  In addition, it is clear that grains can act as a catalyst for chemistry on the surface, producing both simple and complex molecules as part of an interchange between the gas and solid phases \citep{tielens_hagen, herbst_grainsur}.

 \subsection{Observational Chemistry}\label{sec:obschem}
 
 \subsubsection{Gas-Phase Freeze-Out}
 As discussed in \S 3 a sample of low-mass starless cores are now known with well described physical structure as a function of position: N$_{H_2}$(r), n$_{H_2}(r)$, T$_{dust}(r)$, T$_{gas}(r)$.   Sample profiles are given in Fig.~\ref{fig:galli}.  Foremost among these are the two new template objects: L1544 and B68. There are some caveats to the determination of each of these parameters, but the overall structure is well characterized and has led to significant advances in our understanding of the chemistry.   In general there are two factors that influence these gains.  (1) Using the dust mm/submm continuum emission  as a surrogate for the \Htwo\ distribution allows for an examination of molecular abundance variations directly relative to \Htwo .   This is a fact noted by \citet{mezger92} in the high-mass NGC2024 star forming region.  However, the presence of embedded sources can substantially influence the dust emission morphology, potentially leading to different conclusions \citep{chandler96}.  While starless cores do have small thermal gradients (\S 3.2) this particular complication is minimized.  (2) Molecular emission depends on the \Htwo\ density, gas temperature, and on the abundance.  Using sophisticated one-dimensional radiation transfer codes \citep[Monte-Carlo or Accelerated Lambda Iteration;][]{bernes,ali} and the density and temperature {\em structure} as inputs, the abundance profile along the line of sight can be constrained for the first time.

Initial studies found the first direct link between chemistry in the gas and interactions with grain surfaces by simply comparing the estimated CO and CCS column densities to the H$_2$ column estimated from the dust \citep{klv, willacy98, kramer_codep, caselli_codep}.  
This is interpreted as evidence from gas-phase data that the CO molecules are freezing onto grain surfaces.  A comparative study of CO suggest freeze-out dominates  when densities exceed  $\sim 3 \times 10^4$ \cc \citep{bacmann_codep}.   

Spatial mapping of ices within individual cores supports        the interpretation of the  gas-phase observations as the abundance of CO ice is found to significantly increase when densities exceed $\sim 10^5$ \cc\
\citep{klaus_aa}.  Near the core center the abundance is close to the typical gaseous CO abundance of $10^{-4}$, indicating that the majority of CO is frozen out.
  Furthermore, the abundance of water ice is typically $0.5 - 0.9 \times 10^{-4}$ except at the highest densities where the abundance increases \citep{klaus_iau}.    The molecular ices are therefore substantial reservoirs of the available oxygen.

\subsubsection{Selective Freeze-Out}
\label{sec:chemsurv}
\begin{figure}
 \centerline{\psfig{figure=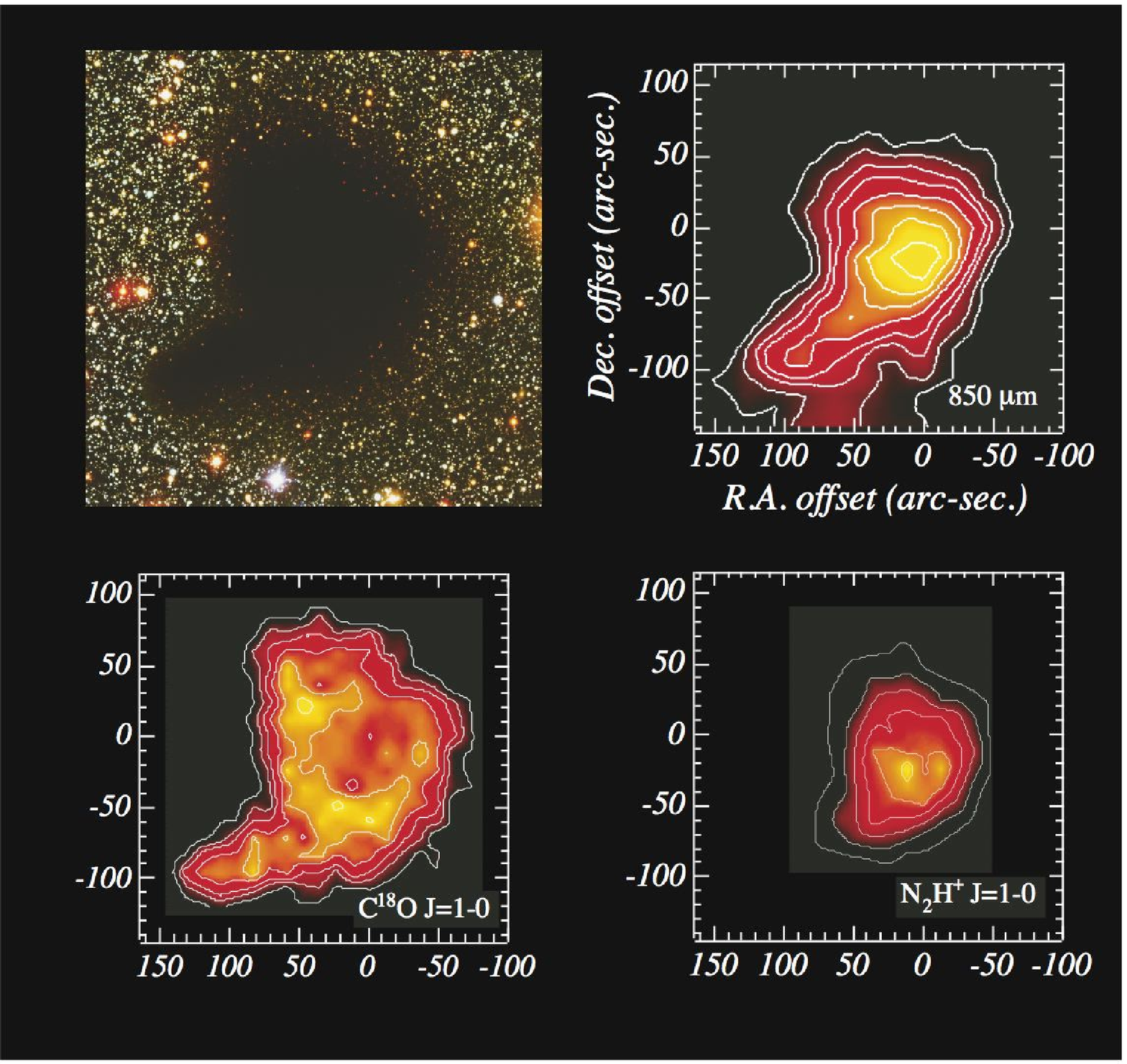,height=5in}}
 \caption{
 A deep optical image of the dark globule
Barnard 68 ({\em top left}; \citealt{alv01}) along
with contour maps of integrated intensity from molecular emission 
lines of N$_2$H$^+$ (contour levels: 0.3--1.8 by 0.3 K km s$^{-1}$), C$^{18}$O (0.2--0.7 by 0.1 K km s$^{-1}$), and 850$\mu$m dust continuum 
emission (10--70 by 10 mJy beam$^{-1}$). Molecular data, with an angular resolution of $\sim 25''$ ,  are from \citet{bergin_b68} and dust emission (angular resolution of 14.5$''$) from \citet{bia03}. 
\label{fig:b68}}
\end{figure}

Molecular surveys reveal further differences,
Fig.~\ref{fig:b68} shows a sample of this work with a comparison of the 850$\mu$m dust continuum distribution in the Barnard 68 starless core with molecular emission traced by \CeiO\ and \NtwoHp .   In this core the \NtwoHp\ emission more closely follows that of the dust than does \CeiO , which appears as a ``semi-ring-like'' structure around the dust and \NtwoHp\ emission maximum \citep{bergin_b68}.   \citet{taf02} surveyed 5 cores in Taurus (including L1544) illustrating general characteristics of core chemistry: carbon-bearing species, represented by CO and CS deplete from the gas while nitrogen-hydrogen bearing molecules, \NtwoHp\ and \NHthree\ trace the core center \citep[see also][]{caselli_codep}.   From the abundance profile it is estimated that the CO and CS abundance traces a large dynamical range with declines of at least 1-2 orders of magnitude in the core centers relative to the lower density core edge, while the abundances of nitrogen molecules either stay constant or decay more slowly.
The interpretation of ``selective'' freeze-out, where molecules exhibit different behavior in terms the response of the chemistry to interactions with grain surfaces,  naturally explains some long-standing issues such as the  emission differences seen between carbon and nitrogen molecules described earlier \citep[compare ][]{swade89, pagani_dep}, and
the larger core sizes and velocity dispersions seen for CO and CS when compared to earlier \NHthree\ data \citep{zhou89, taf02}.  For star formation this isolates the nitrogen hydrides as the key probes of dense gas.
 
\subsubsection{Carbon, Oxygen, and Nitrogen}
Cores with well defined physical properties have clarified additional outstanding chemical issues regarding carbon, oxygen, and nitrogen chemistry.    It is clear that CO represents the major reservoir of carbon in the gas and, along with CO$_2$, in the grain mantle.   \citet{maret_n2} demonstrated that \Ntwo\ is not the dominant carrier of nitrogen in molecular gas.  They suggest that N~I as a  main reservoir of atomic nitrogen 
\citep[see also][]{vdb_araa}; however, at present, it is unclear what is the predominant form of gas-phase nitrogen.   Observations by the {\em Submillimeter Wave Astronomy Satellite} (SWAS) and {\em Odin} also demonstrated that the abundance of ortho-water vapor
and molecular oxygen is well below theoretical expectations throughout the Galaxy \citep{snell_h2o, odin} and in starless cores in particular \citep{bergin_snell}.  
Due to atmospheric absorption, para-water vapor has yet to be observed and
the limits on the water abundance disagree with pure gas-phase chemical theory provided that the ortho/para ratio is not below the equilibrium value at 10 K (o-\HtwoO /p-\HtwoO\ $\sim 0.3$).
The low water vapor abundance is interpreted as the result of water ice formation on grain surfaces.  At grain temperatures below $\sim 110$ K
 (\S 4.3.2) water will remain frozen on grains,
 which indirectly lowers the gas phase atomic oxygen abundance thereby hindering formation of water and molecular oxygen in the gas
\citep{bergin_impact, charnley_isoswas}.
Further analysis suggests that the water vapor emission likely arises from 
the surface photodissociation region where a balance between photodesorption of water ice and photodissociation of water vapor exists (OH should behave in a similar fashion) \citep{melnick_h2opdr}.
The launch of the Herschel Space Observatory, which will be capable of observing both ortho- and para-water will provide definitive information on the water abundance and its potentially defining role in gas-phase freeze-out.
 
\subsection{Chemical Theory}

\subsubsection{Basic Processes}
In the literature the loss of gaseous molecules to the solid phase is often labeled as depletion, which implies a lowering of the gas-phase abundance, or as freeze-out, which is more direct statement of the process.   In practice both terms can be used, but there are some nuances.   While neutral species freeze onto grain surfaces it is not thought that molecular ions behave in a similar fashion.   Grains in dense clouds are believed to carry negative charge \citep{weingartner_draine}.  The ion-negatively charged grain recombination releases a few eV which, if carried by the products, significantly exceeds the  molecule-grain binding energies of $\sim 0.1-0.5$ eV \citep{collings_iau}.   Thus ions can deplete, via, for example, gas phase processes and by pre-cursor molecules (e.g. CO for \HCOp ) adsorbing onto grains,  but do not freeze-out.

Atoms and molecules freeze onto grain surfaces or adsorb with a 
rate of $k_{fo} = \sigma v n_{g}S$ s$^{-1}$.   Here $\sigma$ is the grain cross-section (generally assuming to be an average grain with a 0.1 $\mu$m radius), $v$ is the mean velocity of the Maxwellian distribution of gaseous particles, $n_g$ the space density of grains, and $S$ is the sticking coefficient (i.e. how often a species will remain on the grain upon impact).  
Using relevant parameters for CO at 10 K \citep[assuming $S = 1$ as suggested by laboratory experiments; e.g. ][]{bisschop06} the freeze-out timescale $\tau_{fo} = k_{fo}^{-1} \sim 5 \times 10^9/n_{H_2}$(\cc ) yrs. Once on the grain, molecules can react with any species that is mobile on the grain surface at 10 K, such as atomic hydrogen, or if a species is saturated (e.g. \HtwoO , \NHthree , \COtwo ) it will remain inert on the surface.    
For cold dense gas that is shielded from external radiation the physical processes that can remove or desorb molecules from grain surface are either inactive or can only desorb the most volatile ices \citep[see][for a complete summary of desorption processes]{vdb_araa}.  In cold cores it is thought that energetic cosmic ray impacts can desorb the most volatile ices \citep[e.g. CO but not \HtwoO ;][]{leg85, has93, bringa_cr}. Thus the dominant effect is the adsorption of atoms and molecules onto grain surfaces and any subsequent consequences that result from this interaction.   

\subsubsection{Chemical and Dynamical Timescales} \label{sec:timescale}
\begin{figure}
 \centerline{\psfig{figure=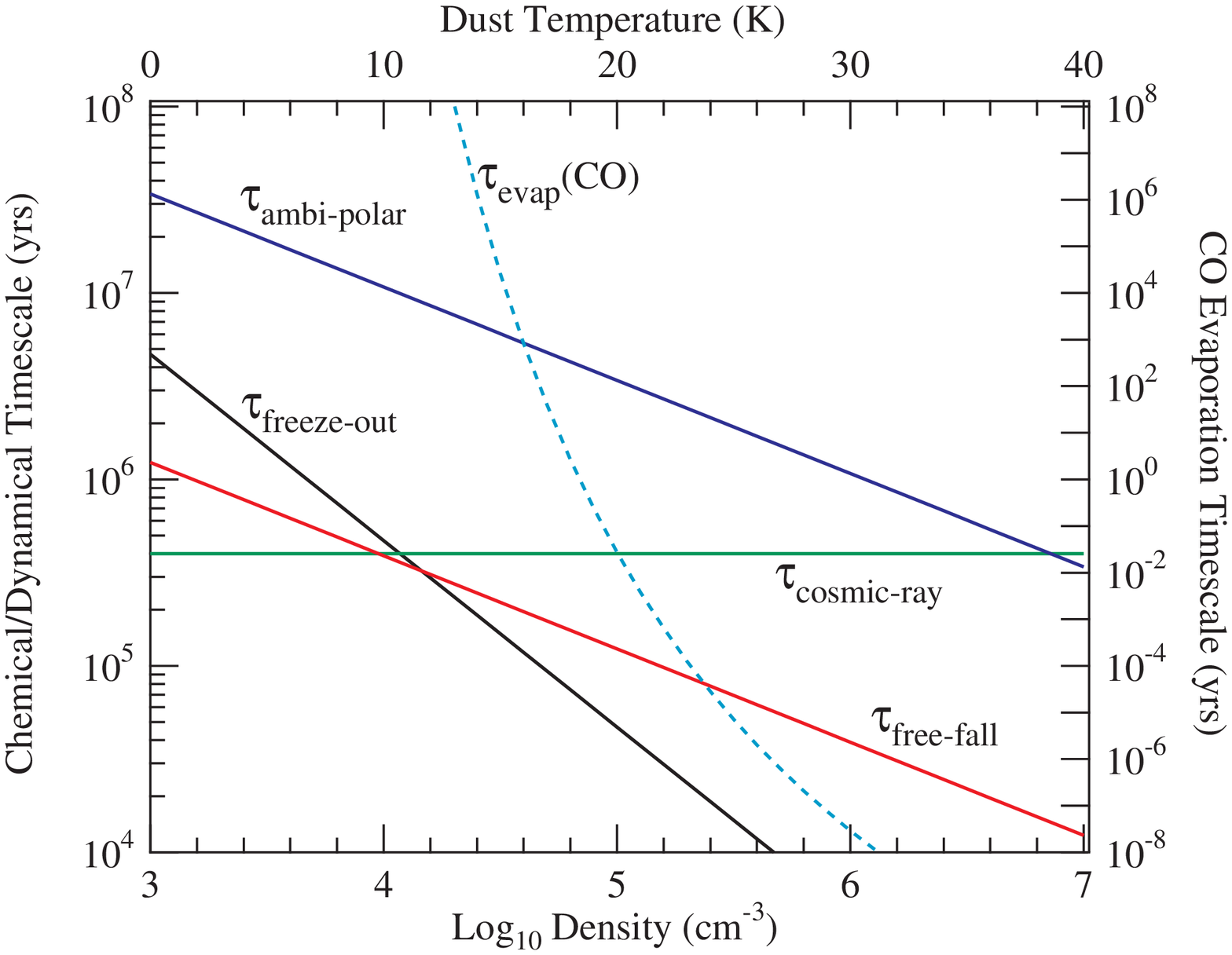,height=4in}}
 \caption{
 Plot of various chemical and dynamical timescales  shown as a function of the molecular hydrogen volume density.   
 Also shown is the density independent CO evaporation timescale as function of the dust temperature (which in the plot is also not a function of the density).  Relevant equations for each quantity are provided in \S~\ref{sec:timescale} and \S~\ref{sec:ions}.  In this plot the cosmic ray timescale is set by the time for any change from an equilibrium state to be reset by cosmic ray powered chemistry.  The ambipolar diffusion term assumes $x_e = 2 \times 10^{-8}(n_{H_2}/10^{5} \rm{cm}^{-3}$)$^{-0.5}$.
 \label{fig:timescales}}
\end{figure}

In Fig.~\ref{fig:timescales} we compare relevant chemical and dynamical timescales.  Representative dynamical timescales are provided for free-fall and ambipolar diffusion, the latter having been suggested as a primary mode of core condensation \citep{shu87}.
The free-fall timescale is $\tau_{ff} = [3\pi/32G\rho]^{0.5} \sim 4 \times 10^7/n_{H_2}^{0.5}$ yrs and the ambipolar diffusion timescale is discussed in 
\S~\ref{sec:ions}.
The gas-phase chemical timescale is set by cosmic rays and has a constant dependence on the density, while freeze-out has linear dependence on the density.  The evaporation rate is given by the Polanyi-Wigner relation \citep{tielens_book}:  

\begin{equation}
\tau_{evap} = \nu_0^{-1} exp(E_b/kT)\;\;(s^{-1})
\end{equation}

\noindent where $\nu_0$ is the vibrational frequency of the molecule in the potential well (typically $\sim 10^{12}$ s$^{-1}$) and $E_b$ is the binding energy.  The evaporation timescale has an exponential dependence on the dust temperature, but critically depends on the binding strength to the grain surface or mantle.    Measured binding energies for CO-CO are $\sim 820$~K \citep{collings_cobind} and $\sim 5800$~K for \HtwoO -\HtwoO\ \citep{fraser_h2obind}, implying evaporation temperatures of $\sim15$ K for CO and $\sim110$ K for \HtwoO\  at \nhtwo\ = 10$^5$ \cc\ and assuming a typical 0.1 $\mu$m grain abundance of $x_{gr} = 10^{-12}$.
Fig.~\ref{fig:timescales} demonstrates a key aspect of starless cores - as cold objects, $T_{dust} \sim T_{gas} \sim 10$ K, evaporation is unimportant, with timescales $> 10^8$ yrs.  Furthermore the freeze-out timescale becomes shorter than the free-fall and ambipolar diffusion timescales as the density increases.
  Thus as the core physically evolves, the 
effects of freeze-out on gaseous species are magnified, and can be used to constrain the mechanism of core formation.   

\subsubsection{Chemical/Dynamical Models}
Combined chemical and dynamical models follow the chemical composition as the core contracts.  Such models generally adopt parameterized fits or a more direct coupling to simple dynamical models  \citep[e.g.][]{larson69, penston69, shu77, basu94}.   Notable chemical/dynamical models are \citet{rawlings_shu}, \citet{ bl97}, \citet{aikawa_lp} and \citet{li02}.
With basic gas-grain microphysics incorporated into time-dependent models \citep[see][]{tielens_book} the chemical structure discussed in \S~\ref{sec:chemsurv} can be explained and, in some cases, was predicted.
 
\begin{figure}
 \centerline{\psfig{figure=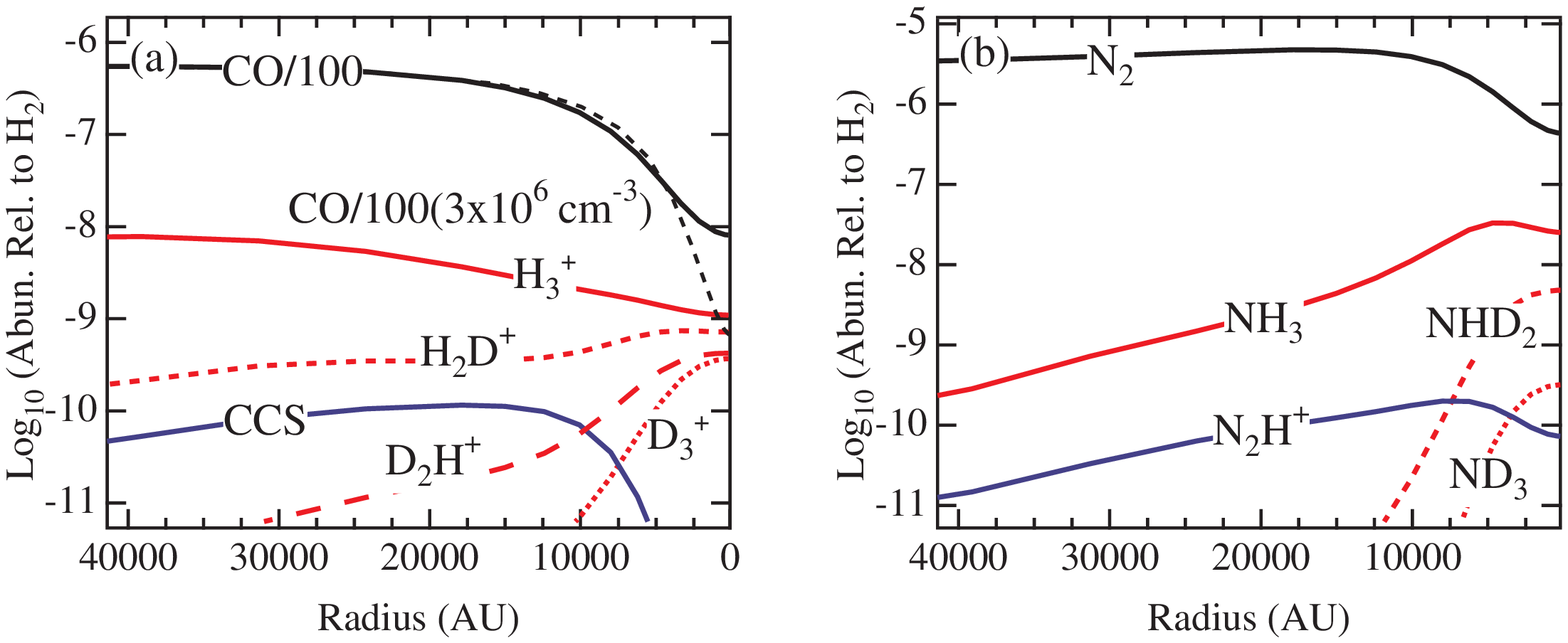,height=2.5in}}
 \caption{
 Plot of chemical abundances as a function of core depth for selected species from a chemical and dynamical model of a contracting Bonnor-Ebert sphere \citep[taken from][]{aikawa_be}.   All species are shown at a timescale where $n_H(center) = 3 \times 10^5$ \cc , excluding CO which also is shown at $n_H(center) = 3 \times 10^6$ \cc .
 \label{fig:chemmod}}
\end{figure}

 Basic elements of the chemistry are illustrated in Fig.~\ref{fig:chemmod}.    During collapse or condensation, the central regions of the core increase in density to levels where  neutrals rapidly freeze onto grains (\nhtwo\ $> 10^5$ \cc ).  This contrasts with the lower density less evolved core edges with undepleted abundances.   As collapse proceeds these effects become amplified (compare the two times provided for CO in 
 Fig.~\ref{fig:chemmod}).
 \citet{bl97} first showed that carbon- and sulphur-bearing species (e.g. CO, CS, CCS) suffer the greatest effect from depletion. In contrast the nitrogen hydride pool is the least affected \citep[see also][]{aikawa_lp}.   At first this was assumed to be due to a lower binding strength of \Ntwo\ (the precursor molecule for both \NtwoHp\ and \NHthree ) to the grain surface than for CO.   This has now been measured to be equivalent to CO \citep{obe05}.
 
  It is now recognized that the observational appearance of selective freeze-out can be separated into first-order chemical effects, the direct freeze-out of neutrals (e.g. CO, CCS, CS, N$_2$), and second-order effects, changes created by the imbalance in the chemistry left primarily by the loss of gaseous carbon monoxide.   
 CO is a major destroyer of molecular ions and its removal from the gas leads to a change in the relative abundance of major charge carriers \citep{rawlings_shu}. Particularly affected is the \Hthreep\ ion, the precursor to molecular ions such as \NtwoHp\ and \HCOp\  that are the respective daughter products of \Ntwo\ and CO.   These key observable ions (\NtwoHp\ and \HCOp ) will also decrease in abundance as the parent molecules freeze-out; however, CO is a major destroyer of \NtwoHp\ and as CO disappears from the gas there is a subsequent increase in abundance of \NtwoHp\ \citep{aikawa_lp, bergin_b68, jesj04}.  \NHthree\ then forms from \NtwoHp\ \citep{geppert_n2hp, aikawa_be}.    Thus the abundance of the nitrogen hydrides is strongly enhanced and these species are probing the gas where CO (and other carbon-bearing species) are depleting.  An important point here is that the abundance of both \NHthree\ and \NtwoHp\ are several orders of magnitude below the abundance of molecular nitrogen.  \Ntwo\ is freezing onto grains at the same level as CO.   Provided the \Ntwo\ abundance does not deplete to levels below \NtwoHp\ and ammonia, then these  tracers can still form and emit from layers where commonly observed tracers of dense regions have significantly reduced abundances.

Combined models  reproduce not
only the pattern of depletions, but also the column densities observed
in L1544 \citep{aikawa_lp, li02}.    In general models obtain best agreement with  collapse timescales much shorter than would proceed from ambipolar diffusion starting from a magnetically subcritical state \citep[e.g.][]{walmsley_dep}; however there is some disagreement \citep{li02}.
A clear point is that evolutionary timescales at densities near $\sim 10^{4}$
 \cc\ cannot be much longer than $\sim 0.5 - 1$ Myr  as significant CO freeze-out would be produced at densities below where it is observed \citep{flower_dep, tafalla06}.    Moreover, collapse cannot be too fast ($>$ 0.2 km s$^{-1}$ at $\tau_{dyn}
  \lesssim 5 \times 10^5$ yrs)\footnote{The typical non-thermal linewidth in starless cores (e.g. L1498, L1517B, B68, ...)  is 0.2 km/s. 
  If this  represents a 
purely inward motion, it would correspond to an  infall speed of
0.1 km/s at most. With a typical core radius is 0.05 pc, this 
implies a typical contraction time of 0.5 Myr.} or  there will be less depletion and emission linewidths would be too broad \citep{myers83}.
 
Current work has greatly benefited from laboratory work on low temperature gas phase reactions, grain binding energies, and on surface catalysis \citep{drecomb_rev, collings_iau, nag05, bisschop06, liv03, katz_h2}.  Moreover the recent re-calculation and extension of collisional rates of astrophysically important molecules is of great use and deserves mention  
\citep{dubernet_h2o, flower_co, daniel_n2hp}.
 However, large questions remain regarding the nature of the grain surface catalysis in this cold environment and the strength of the mechanisms that could return molecules to the gas  \citep[e.g.][]{leg85, bringa_cr, shen04}.
It is also unclear how complex molecules form in the gas-phase.  For example \CHthreeOH\  is detected in cold clouds, but currently has no known gas-phase formation route.   An examination of the extent of the complex chemistry in cores with well defined properties is needed and will likely lead to gains in our understanding of interstellar organic chemistry \citep[see][]{hir04}.

One clear positive statement is our knowledge of the chemistry and the time sensitive nature of the gas-grain interaction is such that we now have the methods to place relative ages between cores within a given sample \citep{hirota_l1521e, tafalla_l1521e}.   This should aid in our analysis of the core/star formation process.  Moreover,
the combination of chemical/dynamical models with radiative transfer \citep[see][as an example]{lbe04} offers the best opportunity to take these models to the next level and truly begin to test star formation theory.
 

\subsection{Deuterium Chemistry and the Ionization Fraction}

\subsubsection{Deuterium Fractionation}
Another second-order effect induced by the CO freeze-out is a sharp rise in the level of deuterium fractionation.   Enhancements of deuterium bearing molecules relative to the hydrogen counterparts has been known for some time \citep{guelin82}.   In cold clouds enrichments of 2-3 orders of magnitude are observed above the atomic hydrogen value of (D/H) $\geq (2.3 \pm 0.2) \times 10^{-5}$ estimated within 1 kpc of the Sun \citep{linsky_d}.   Because of the lower zero point energy of deuterium bonds compared to bonds with hydrogen, ion-molecule reactions in the dense ISM are thought to be the mechanism responsible for these enrichments \citep{mbh}.    Deuterium enrichments can also be created via reactions on the surfaces of dust grains \citep{tielens83, nag05}, but in these cold cores freeze-out is believed to dominate as opposed to sublimation (e.g. freeze-out dominates the chemistry of the volatile CO molecule).

In the cold dense ISM, deuterium chemistry is driven by the following reaction:

\begin{equation} \label{eq:h2dpform}
\rm{H}_3^+ + HD \leftrightarrow  \rm{H}_2D^+ + H_2 + 230 K.
\end{equation}
 
 \noindent  The forward reaction  is slightly exothermic favoring the production of \HtwoDp\ at 10 K, enriching the [D]/[H] ratio in the species that lie at the heart of interstellar ion-molecule chemistry \citep{mbh}\footnote{Reaction~\ref{eq:h2dpform} has been measured in the lab at low temperatures finding that there is an additional dependence on the ortho/para ratio of \Htwo\ \citep{gerlich_d}.}.  These enrichments are then passed down the reaction chains to species such as \DCOp , DCN, HDCO, and others.

Pure gas-phase models without freeze-out cannot produce significant quantities of doubly \citep[NHD$_2$;][]{roueff_nd2h} and triply deuterated ammonia \citep[ND$_3$;][]{lis_nd3} as observed in starless cores.
 This motivated a re-examination of the basic deuterium chemistry.  The primary advance in our understanding is two-fold: (1) deuteration reactions do not stop with \HtwoDp , rather they continue towards the formation of both \DtwoHp\ and \Dthreep , via a similar reaction sequence \citep{phillips_vastel, roberts_d, walmsley_dep}:

\begin{equation}
 \rm{H}_2D^+ + HD \leftrightarrow  \rm{D}_2H^+ + H_2 + 180 K,
\end{equation}

\begin{equation}
  \rm{H}_2D^+ +  HD \leftrightarrow D_3^+ + H_2  + 230 K.
\end{equation}

\begin{figure}
 \centerline{\psfig{figure=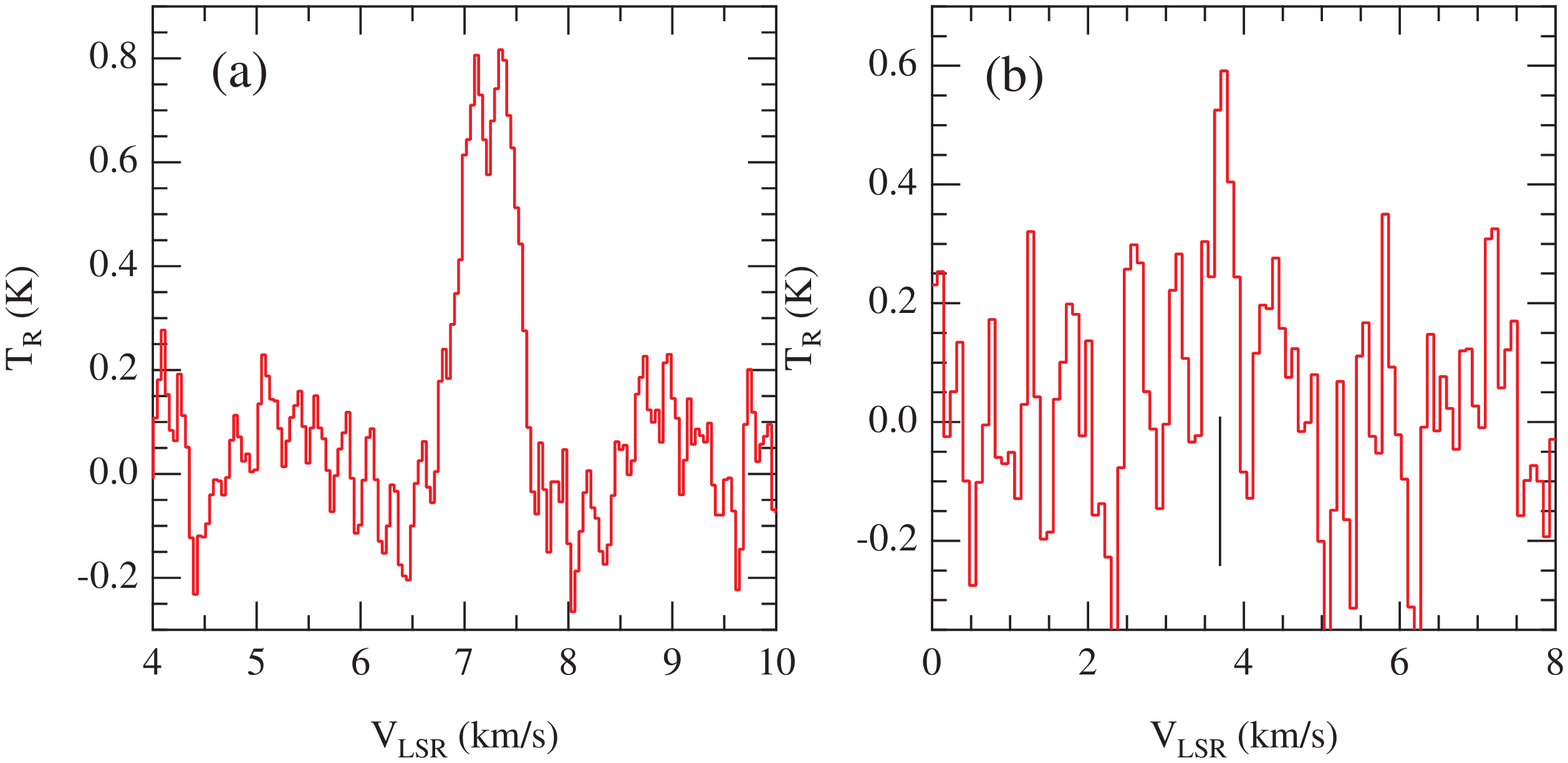,height=2.5in}}
 \caption{
 (a) Detection of \HtwoDp\ emission in L1544 \citep{caselli_h2dp, vdt_h2dp}.   (b) 4.4$\sigma$ detection of \DtwoHp\ in the starless core IRAS16293e \citep{vastel_d2hp}.
 \label{fig:D}}
\end{figure}

\noindent (2) The freeze-out of CO, which is primary destroyer of both \Hthreep\ and \HtwoDp , increases the rate of the  gas phase fractionation reactions 
\citep[see deuterium species in Fig.~\ref{fig:chemmod};][]{svv99, aikawa_lp, bacmann_d}. This inference is strongly supported by the detection of \HtwoDp\ and \DtwoHp\ in starless cores
\citep{caselli_h2dp, vastel_d2hp, vdt_h2dp}, as illustrated in Fig.~\ref{fig:D}.  

 The creation of multiply deuterated ions leads to the production of high deuteration levels similar to those observed.  An additional by-product is an enhanced atomic (D/H) ratio which will accrete onto grains and be available for surface reactions \citep{tielens83, cha97, stantcheva03, aikawa_be} creating molecules rich in deuterium to be exposed when ices sublimate off warm grains as the result of stellar birth.   However, the extent of the surface deuteration opposed to the gas phase needs to be further explored.   It is worth emphasizing that the observed form of \HtwoDp\ in interstellar clouds is ortho-\HtwoDp\ and models show that the \HtwoDp\ ortho/para ratio ($o/p$) will change with density and is a sensitive function of the molecular hydrogen ortho/para ratio \citep{flo06}.   If the  H$_2$ $o/p$ ratio is not in equilibrium at the low $T_g < 20$ K (where $o/p \ll 1$) and instead is near 3:1, then deuterium fractionation will not work \citep{pfm91, gerlich_d}.   Observations from the Spitzer Space Telescope suggest that the H$_2$ $o/p$ ratio is in equilibrium in cold pre-shock gas \citep{neu06}, a fact consistent with the observed fractionation.  Herschel and SOFIA will be needed to  constrain the H$_2$ $o/p$ ratio via observations of ortho- and para-\HtwoDp\ in starless cores.

\subsubsection{Ionization Fraction} 
\label{sec:ions} For subcritical starless cores (where the magnetic field supports the cloud against gravity, \ref{sec:cloud_bfield}) the evolutionary timescale is set by ambipolar diffusion and the ion fraction is a critical factor in determining the magnetic flux leakage.   \citet{mckee89} discussed the various charge carriers and recombination rates deriving a basic expression that has wide use: $x_e = 1.3 \times 10^{-5}n($\Htwo )$^{-0.5}$.
The effects of freeze-out can alter both the pre-factor and the exponent of the density dependence \citep{bl97, caselli_l1544ion, walmsley_dep}, but see also \citet{ciolek_elec}.

 Past studies of the ion fraction in dense cores focused on the \DCOp /\HCOp\ ratio, that was predicted to be a sensitive function of the electron abundance.
 Furthermore, it was thought that \HCOp\ was a major charge carrier throughout the core \citep[e.g.][]{caselli_ion, williams_ions}.  These studies found line of sight average ion fractions of $x_e \sim 10^{-7}$ (relative to H$_2$), implying weak coupling to the magnetic field.   One variable assumption in models is the presence or absence of metal ions (e.g. Fe$^+$, Mg$^+$, ...), which have slower recombination timescales than molecular ions.  If present, metals could dominate the electron fraction over molecular ions ultimately produced by cosmic ray impacts.  Currently detailed models imply no significant contribution from heavy metal ions to the electron fraction \citep{caselli_l1544ion, maret_n2}.   

\begin{figure}
 \centerline{\psfig{figure=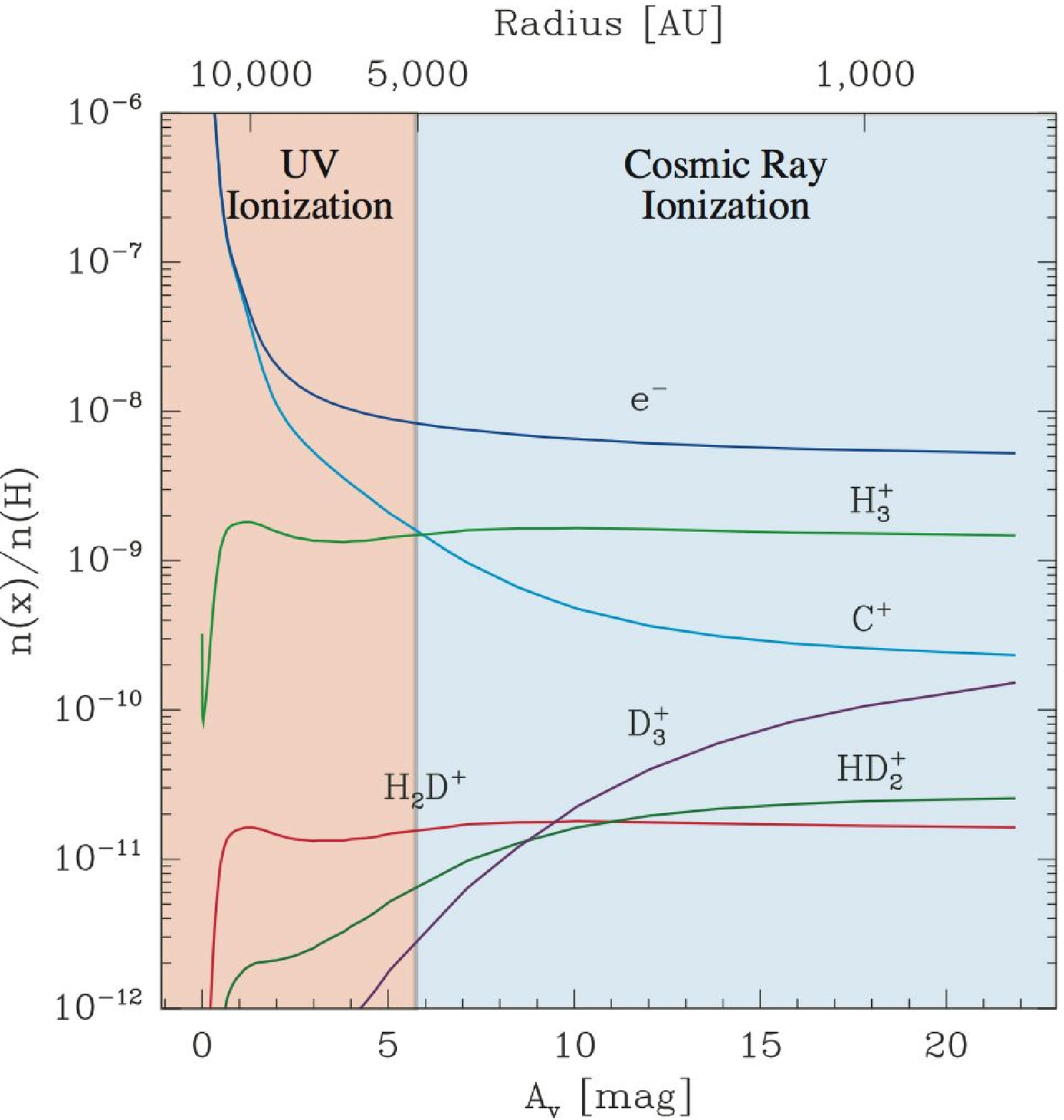,height=4.5in}}
 \caption{Derived electron fraction and major ion abundances in the Barnard~68 core.  Abundances are relative to H nuclei.  Zones where the ionization is dominated by UV and Cosmic Ray are delineated (effectively where C$^+$ is no longer the most abundant ion).  
 In this plot H$_3^+$ is the main ion, carrying $\sim 20$\% of the charge with the remainder contained in more complex ions.   Taken from Maret \& Bergin (2007), in prep.   Some uncertainties in these calculations include the gas-grain model (i.e. the assumed binding strengths and desorption mechanisms) and the cosmic-ray ionization rate.
 \label{fig:electron}}
\end{figure}

Due to the freeze-out of neutrals, \HCOp\ does not trace the core center and major charge carriers in the core center are likely to be a combination of \Hthreep , \HtwoDp , \DtwoHp , and \Dthreep .  An example of the ``typical'' ionization structure is given in Fig.~\ref{fig:electron}.   When illuminated by the standard interstellar radiation field or below, the ionization is high at core edges (\av\ $<$ 1--2 mag) due to CO photodissociation and ionization of carbon.  Deeper in the cloud the regime of cosmic ray ionization leads to an ion fraction that decays with increasing density.
 For the densest (\nhtwo\ $> 10^6$ \cc ) most evolved cores the ultimate expectation is that \Dthreep\ is the dominant ion 
\citep{roberts_d, walmsley_dep}.
Observations and models now suggest that in at least some objects the ion fraction in the core center is quite low, $x_e \sim 3 \times 10^{-9}$, implying that the gas is locally decoupled from the field \citep[i.e. the timescale for ambipolar diffusion is comparable to or less than the free-fall time,][]{caselli_l1544ion, vdt_h2dp}.

 \citet{walmsley_dep} and \citet{flower_dep} discussed ambipolar diffusion timescales in the context of depletion, providing detailed expressions:

\begin{equation}
\tau_{ad} \sim \frac{2}{\pi Gm_{\rm H_2}^2}\sum_i \frac{n_i}{n_{\rm H_2}} \mu_{in}<\sigma v>_{in}\;\;{\rm yr}
\end{equation}
 
\noindent where G is the gravitational constant, $i$ implies a summation of ionic species, $\mu_{in} = m_i m_{H_2}/(m_i + m_{H_2})$ is the reduced mass with H$_2$ the neutral collision partner.  The rate coefficient for momentum transfer between ions and neutrals is assumed to be $< \sigma v>_{in} = \pi e (\alpha/\mu_{in})^{0.5}$, with $\alpha$ as the polarizability of \Htwo\ \citep{osterbrock61}.   If \HCOp\ is the dominant ion then $\tau_{ad} \sim 1.7 \times 10^{14}x_e$.  An essential feature is that the timescale is proportional to the square root of the reduced mass implying that the ionic composition also influences the diffusion of magnetic flux (the timescale is $\sim 60$\% lower if \Hthreep\ dominates and 25\% for \Dthreep ).  These works highlighted the potential of negatively charged grains to alter the degree of ionization and also as limiting factors in deuterium fractionation, which can be influenced by the changing size distribution as the result of coagulation.  

\subsection{Key Tracers and the Possibility of Complete Heavy Element Depletion}
 
 Starless cores have revealed a dynamic and evolving chemical structure that highlights key tracers that probe different depths and evolutionary times.   One open possibility is that all heavy elements are depleted in core centers prior to stellar birth.  This has important implications for our ability to  study the process because the heavy element species (e.g. CO, CS, \NtwoHp , ...) have permanent dipole moments and can be observed via pure rotational transitions in regions of the atmosphere with good transmission via heterodyne techniques at mm wavelengths.    If the timescales of star formation are such that heavy element depletion occurs prior to the formation of a point source then our ability to study the entire process will be severely hampered, leaving only hydrogen/deuterium molecules with dipole moments as probes (\HtwoDp , \DtwoHp ).
 
 \begin{figure}
 \centerline{\psfig{figure=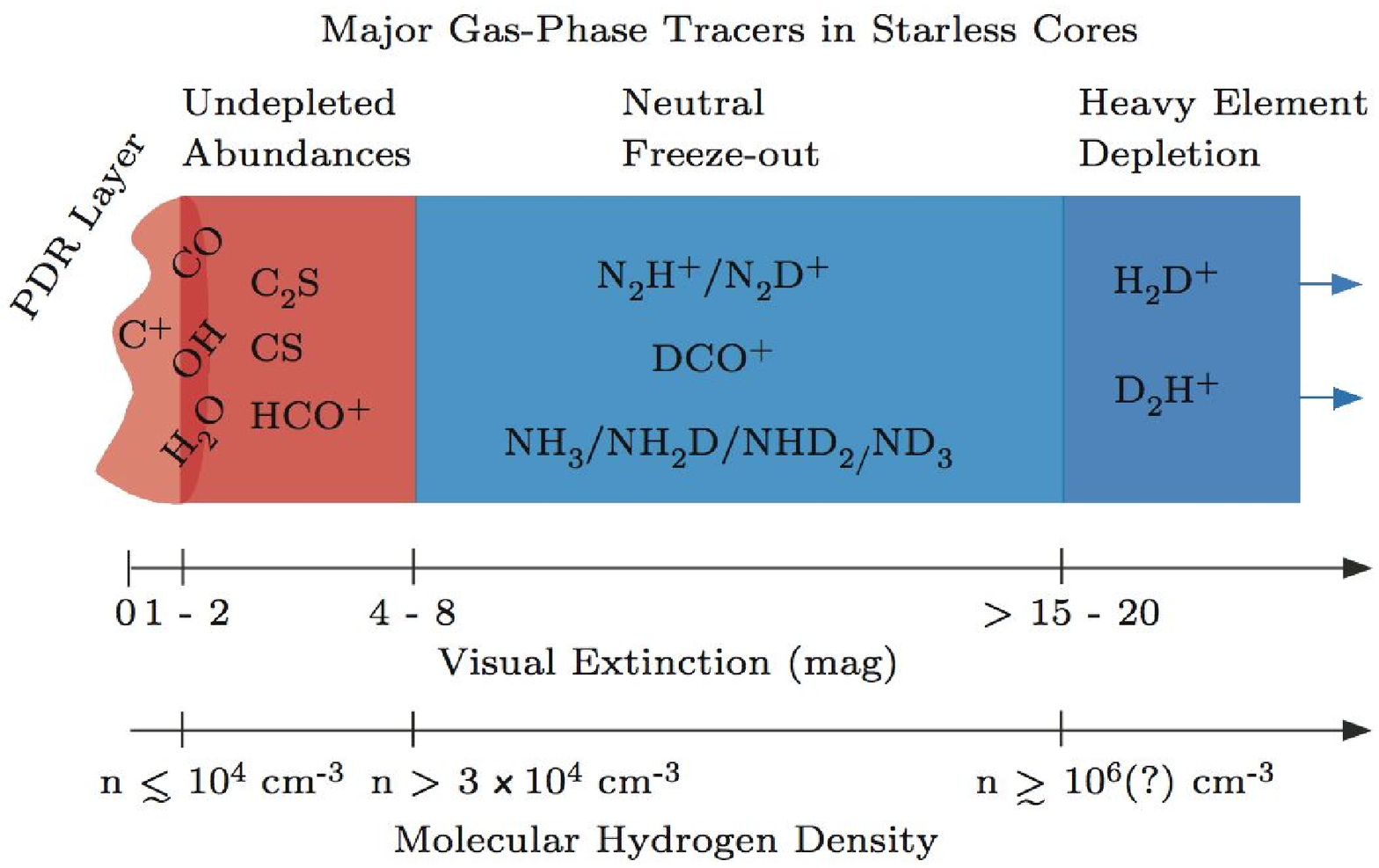,height=3.5in}}
 \caption{
Schematic summary of the major gas-phase probes of starless cores as a function of depth and density.   The scales provided are approximate and were estimated from Barnard 68.   In this schematic the temperature is assumed to be low ($< 15$ K) below A$_V \sim 2$ mag.
 \label{fig:schematic}}
\end{figure}

 At present the evidence for total heavy element depletion is inconclusive and focused on the question of whether \NtwoHp\ is seen to deplete.  This molecule is key as it also is the likely formation route of \NHthree , and if it depletes then ammonia should follow suit.
 \citet{bergin_b68} and \citet{maret_n2} find that \NtwoHp\ does drop in abundance towards the center of the B68 core. 
  In contrast, the \NtwoHp\ and \NHthree\ emission in L1544 when observed in high resolution appears to coincide with the continuum emission peak \citep[][Crapsi et al. 2006, in prep.]{wil99}.   Thus in the two best studied sources we find contradictory solutions and total heavy elemental depletion remains an intriguing possibility.
  
  Given the gains in our understanding of the chemistry and how it changes with core evolution, in Fig.~\ref{fig:schematic} we provide a schematic to summarize our current knowledge.   For less evolved sources ($n_{r=0} <  few \times 10^4$ \cc ) the effects of depletion are likely minimized and commonly observed molecules such as CO, \CeiO , CS, and \HCOp , remain viable tracers.  For evolved sources ($n_{r=0} > few \times  10^4$ \cc )
  species such as \NtwoHp , \NHthree , \NtwoDp , and \DCOp\ appear to be the best molecular probes of core gas, perhaps even to the dense center.  Regardless of whether total heavy element depletion has occurred, both \HtwoDp\ and \DtwoHp\ are unambiguous tracers of the densest gas.  In contrast, emission from  \CeiO , CS, \HtwoO , and \HCOp\ preferentially sample the  core edge.

\section{INFRARED DARK CLOUDS}

\subsection{Background}

\begin{figure}
\centerline{\psfig{figure=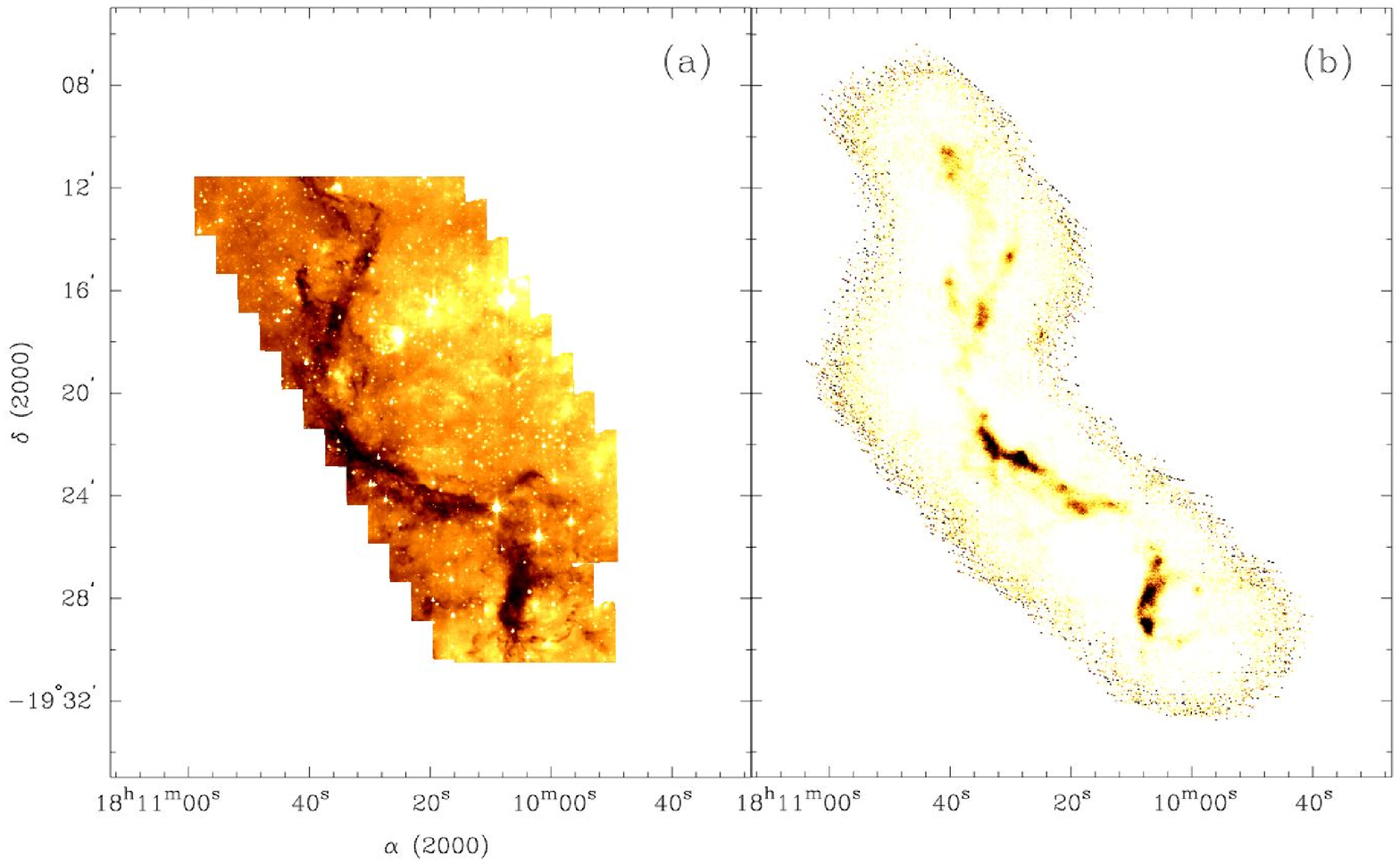,angle=-90,height=5in}}
\caption{G11.11-0.12 infrared dark cloud as seen by (a) Spitzer IRAC band 
4 and  (b)  850 $\mu$m submillimeter continuum emission \citep{johnstone_g11}.  
In the Spitzer image the bright infrared background is provided by PAH emission 
within the 8$\mu$m band and the cloud appears in absorption.  In contrast, the 
submm continuum emission highlights the dense regions  seen in absorption in 
the mid-infrared.\label{fig:irdc}}
\end{figure}

Infrared dark clouds (IRDC) were discovered serendipitously in mid-infrared surveys of the galactic plane by the Infrared Space Observatory's ISOGAL survey \citep{perault_isogal}\footnote{ISOGAL consisted of a 7 and 15 $\mu$m survey of 12 degrees in the plane interior to $| l | = 45^{\circ}$.}
 and the Midcourse Space Experiment  \citep[MSX;][]{egan_msx}\footnote{MSX surveyed from $l = 91-269^{\circ}$ with $b \pm 5^{\circ}$ in 5 filters sampling between 4 -- 25 $\mu$m.  IR dark clouds are best seen in MSX Band A images.}.   Both surveys revealed a population of objects that were dark at mid-infrared wavelengths with opacities ranging from 1--4 at 8 and 15$\mu$m \citep{carey_msx, hennebelle_isogal}.       The objects appear to be quite filamentary and comparison with IRAS suggests that the objects were dark from 7 -- 100 $\mu m$ and ``represent a population of heretofore unrecognized population of cold, dense, isolated clouds'' \citep{egan_msx, hennebelle_isogal}. 
 A compilation of MSX dark objects finds over 10,000 sources with kinematic distances between 1--8 pc with a distribution that peaks towards prominent star forming regions, spiral arm tangents, and the 5 kpc Galactic Molecular Ring \citep{simon_msxcat}.   Fig.~\ref{fig:irdc} presents one example showing the G11.11-0.12 IRDC as seen by the Spitzer Space Telescope and in sub-millimeter continuum emission. 
 
Much of the interest generated by this discovery has focused on the formation of high-mass stars and stellar clusters.  This review demonstrates many of the gains in our knowledge of the formation of low-mass stars through the isolation of starless cores as a unique sample to study the initial conditions of star formation.   In the case of clustered star formation the nearest region is Orion at $\sim 500$ pc and it is unclear how representative Orion is when compared to star formation in the inner galaxy, where most of the molecular mass resides.   IRDC's provide the capability to examine the early stages of star cluster formation beyond the local solar neighborhood.  
The primary issue for massive star formation is that the timescales are much faster when opposed to the formation of solar mass stars.  Thus there will be fewer pre-stellar massive objects present at any given time to isolate.  Moreover, since massive stars are born in more distant clouds with stellar clusters the chances of confusion are greater \citep[for a nice review, see][]{garay_lizano}.   The discovery of infrared dark clouds provides a new -- greatly expanded -- sample to 
search for the precursors to massive stars.   Without constraining observations, theories of massive star formation are divided between models that suggest that high mass stars form  by accretion of an extended envelope or competitive accretion of cluster stars for remaining unbound gas, with stars located near the central potential obtaining higher masses \citep{krumholz06, bonnell06}.     A discussion of the theoretical  issues is beyond the scope of this review and the reader is referred to the cited references and the contribution by Yorke \& Zinnecker in this volume.

\subsection{Physical Properties and Implications}

The properties of infrared dark clouds have been estimated using a variety of
techniques that have been outlined in \S 3 and \citet{evans_araa}.    From
observations of CO, \NHthree, \CHthreeCCH , and  \HtwoCO\ temperatures are
estimated to be below 20 K with densities in excess of 10$^5$ \cc\
\citep{carey_msx, teyssier02, pillai06}.    Surveys of mm/submm continuum
emission and \NtwoHp\ constrain typical core masses of 100-1000 M$_{\odot}$ and
sizes of $R \sim 0.3-0.7$ pc which are comparable to those of molecular cores 
containing embedded high-mass protostellar objects.
 \citep{sridharan_irdc, ragan_irdc, rathborne_irdc}.    The cores are dominated
by turbulence with velocity linewidths of $\sim 2-3$ km s$^{-1}$. These
linewidths are systematically less than seen towards cores with embedded
protostars, which is similar to the dependence observed in local clouds
\citep{sridharan_irdc, ragan_irdc}.    The core mass spectrum has a powerlaw
slope of 2.1$\pm 0.4$ \citep{rathborne_irdc}, which is comparable to the
Salpeter IMF;  however, it is likely that these objects will fragment on scales
below the observed resolution (11$''$), and this value is also consistent with
the CO mass spectrum of molecular clouds (\S 2.3.4).
Based on current evidence, it is clear that these sources likely represent the
birth sites of stellar clusters and that a sub-sample must be forming massive
stars.  
 
 The implications of IRDC's for galactic star formation have been discussed by
\citet{rathborne_irdc} who estimate that the star formation rate in these
objects is $\sim 2$ M$_{\odot}$/yr.  This is comparable to the total Galactic
rate of 3--5 M$_{\odot}$/yr \citep{prantzos_sfr} and hints that this population
is responsible for current star formation in the Milky Way.  
The full promise of these objects will be realized as Spitzer galactic plane
surveys (GLIMPSE and MIPSGAL) are analyzed and 
 sensitive high resolution instruments are brought to bear (e.g. CARMA, PdB,
eVLA, ALMA)  to provide the kind of spatial resolution that is currently
available towards the Orion Molecular Cloud.  Such observations, when combined
with the techniques developed for local starless clouds described in \S 3--4,
will ultimately aid in constraining the fragmentation of molecular clouds and
the eventual formation of stellar clusters including massive stars.
 
 \section{SUMMARY AND FUTURE PROSPECTUS}
 
Cold Dark Clouds are a fundamental component of the galactic interstellar medium.  
They represent the nearest sample of star-forming molecular material, and as such,
they play a key role in unraveling the still mysterious process via which stars and 
planetary systems are born. As this review has shown, the study of dark clouds has 
seen some significant progress but still 
contains significant areas of uncertainty.

On the large parsec-scale, high resolution maps of optical and infrared extinction  
have provided a needed ``big picture'' look at cloud structure, and have determined 
that a significant amount of material lies at low extinction (\av\ $< 3-5$). 
Complementing these data are maps of molecular emission where, aided by improved 
sensitivity and mm-wave heterodyne arrays, we can probe deeper using \thCO\ and 
\CeiO\ than allowed by optically thick $^{12}$CO lines. These data have highlighted 
the filamentary and irregular structure of molecular clouds, suggestive of a 
transient nature.  The velocity structure, on the other hand, tells 
a tale of the dominance of 
turbulence, and sophisticated techniques to characterize gas motions have 
constrained the spectrum of turbulence allowing us to move one step beyond 
Larson's Laws.  Magneto-hydrodynamic models have matured to the point where 
they can produce realistic turbulence, and numerous simulations have shown that 
the decay of MHD turbulence is rapid and unavoidable without a continuous 
source of replenishment. This raises an important challenge to established 
theories of long-lived clouds ($>$ 10 Myr).   
In addition to turbulence, static magnetic fields are a 
potentially important ingredient of cloud physics. Measurements of this
field show that in some instances the polarization vectors follow cloud structure, 
but the measurement and interpretation are complicated by various effects.  
 
  In regions where turbulence has dissipated, dense self-gravitating cores form. 
By mapping their molecular line and dust continuum emission,
we have found that cores are characterized by a 
central region with near constant density and gas temperature 
(to within a factor of 2), together with subsonic motions. 
Bonnor-Ebert spheres often provide accurate fits to the observed emission, 
and although they represent a simplification of the core physics,
they should be considered the best available starting point for numerical 
models of core evolution. 
As in the case of the large-scale, the role played by the magnetic field is 
still uncertain.  In addition, detailed radiation transfer analysis suggest 
the presence 
of motions along the line of sight.  These motions may have an 
inward component that is associated with core condensation, but often contain 
additional contributions of an unknown origin. Core-envelope velocity 
differences, pulsation, or rotation may be responsible for them.
 
Knowledge of core physical structure has revealed the complex chemical 
composition of starless cores.  Observations demonstrate that chemical 
inhomogeneities are an inherent dense cold core characteristic that needs 
to be considered when interpreting molecular line data.   The combination 
of sophisticated radiation transfer codes and realistic core physical structures 
allows, for the first time, the investigation of line of sight chemical gradients. 
As a result, it has been found that
gas-grain interactions dominate the chemistry of cores, and 
this has a number of consequences.  The primary effect is the freeze-out 
onto grains of important gaseous species, such as CO.   This removal of CO 
from the gas phase leads to the production of new species, like 
\NtwoHp\ and \NHthree , in the denser gas where these effects are dominant. 
In addition, the deuterium fractionation of the core 
is greatly enhanced, as multiply-deuterated molecules are produced via 
reactions that begin with \HtwoDp , \DtwoHp , and \Dthreep . Thanks to this
new understanding of the core chemistry, the observed
abundance profiles are beginning to be successfully reproduced by models 
that combine chemistry and dynamics of core contraction. Using these models
we can now isolate molecular probes that target specific 
regions along the line of sight. For example, two classical probes
of core structure, CO and CS, are now understood to be only tracers of the low density 
cloud surface.   Deeper layers of the core are traced by \NtwoHp\ and \NHthree\ (along 
with their deuterated daughter products), while the center is 
unambiguously traced by \HtwoDp\ and \DtwoHp .   These chemical signatures 
change as a  function of core evolution and may help to trace the history 
of gas contraction. 

Within this review we have highlighted areas where progress is needed, and 
here we discuss some areas where breakthroughs could occur using future 
instruments and advance techniques.    For the most part we are concentrating 
on issues that can be addressed via observations. 

\begin{itemize}

\item There is considerable need to connect the molecular cloud component 
with the atomic diffuse interstellar medium.  This may illuminate the 
question of cloud formation, cloud destruction, and the origin of the 
turbulence in molecular clouds, and it will require the combination of 
mapping observations of H~I, extinction, [C~II], and CO emission.    

\item To address the issue of cloud fragmentation and the potential breakdown 
in hierarchical structure, it is required that clouds are mapped 
with a uniform tracer from 
the large to the small scales.  The most likely tracer for this work
would be the dust continuum emission, if observations retain sensitivity to 
large scale structure, and SCUBA-2 offers the best opportunity in this regard.   
Molecular tracers will provide needed velocity information. Unfortunately, 
a wide variety of disparate techniques is currently used to analyze the rich 
information contained in 3-D data cubes. A common description of cloud
structure is clearly needed 
both to inter-compare observations of different clouds and to test numerical 
simulations.

\item  The transition from turbulent cloud to the quiescent core regime 
is a critical element in the evolution towards star formation.  In this 
transition, turbulence dissipates and the cores acquire the flattened 
density profiles reminiscent of Bonnor-Ebert spheres.  Whether the 
observed subsonic motions represent a key part of this transition 
should be explored. For this, an improved understanding of the chemistry 
should help in the selection of a number of  molecular tracers to probe 
different layers.  To track turbulent dissipation, a number of cores spanning a 
range of evolutionary states needs to be isolated and characterized.  Millimeter 
observations with existing and future radio telescopes, preferably 
with large format heterodyne receiver arrays (e.g. SEQUOIA on LMT), 
offer the potential to examine these issues.

\item A clear advance results from the ability to combine 
measurements of the dust column
spanning a range of wavelengths from near-IR absorption to submm emission to
examine the dust properties, in particular the mass opacity coefficient.   
In parallel to understanding how the gas changes with contraction there is an open
question as to how dust properties change  as the mantles grow and grains
coagulate. Herschel and SOFIA observations should aid to disentangle  the coupled
effects of dust temperature and emissivity on dust emission/absorption.

\item Models and observations have mapped a chemical sequence that can track the
relative evolutionary status of condensing cores. Observations can be
used to provide a rough indication of the evolutionary state, but improved
theoretical models are still needed to provide core studies
with an accurate chemical clock to test core condensation theories.
Further testing of models against observables (e.g. integrated
emission and line width) also requires accurate collisional rates, molecular rest
frequencies, measurements of low temperature gas phase reaction rates, and
studies of the gas-grain interaction in the laboratory.
 
\item  It is not clear yet whether the centers of cores are completely
depleted in key elements (C, O, and N) preceding the onset  of gravitational
collapse, and this has important implications in the ongoing search for
star-forming infall.
Herschel, SOFIA, and especially ALMA will be important instruments 
to clarify this issue.

\item The emerging field of infrared dark clouds holds the promise of
unlocking the secrets of massive star formation, which is an area
with considerable theoretical controversy.  At present it is not certain how
IRDC's fit into our picture of Giant Molecular Clouds.   Are these sources the
dense centers of clouds in the Galactic molecular ring, and if so, how do they
compare in properties (mass spectrum, size, clumping/fragmentation, turbulence)
to the local population of massive star-forming regions exemplified by Orion?
Systematic studies with high resolution ($< 4''$) are needed to bring 
the study of these objecst to the level of detail that
currently exists for local clouds. eVLA and ALMA will be important in this regard.

\end{itemize}

\noindent ACKNOWLEDGMENTS

We thank Y Aikawa, P Caselli, M Heyer, P Myers, M Walmsley, and J Williams for reading
portions of the manuscript and providing useful comments. E van Dishoeck gave the 
manuscript a thorough reading and suggested a number of corrections that are greatly
appreciated. We are grateful to Y Aikawa, A Bacmann, P Caselli, D Galli, P Goldsmith, 
D Johnstone, C Lada, C Rom\'{a}n-Z\'{u}\~{n}iga, S Schnee,  F van der Tak, and 
C Vastel for supplying figures or material for figures prior to or post-publication. 
We are also grateful to members of the c2d team who provided information well in 
advance of publication, in particular, to N Evans, T Bourke, M Enoch, and M Dunham. 
EAB's work has been supported by the National Science Foundation 
under Grant No. 0335207, and MT's work has been partly supported by project
AYA 2003-07584 from the Spanish Ministerio de Educaci\'on y Ciencia.

\end{document}